\documentclass[11pt,a4paper]{article}

\usepackage{caption}
\usepackage{subcaption}
\usepackage{jcappub}
\usepackage{bm}
\usepackage{epsfig}
\usepackage{bbold}
\usepackage{graphicx}
\usepackage{amsmath}
\usepackage{amssymb}
\usepackage{amsbsy}
\usepackage{color}
\usepackage{slashed}
\usepackage{afterpage}
\usepackage{psfrag}
\usepackage{hyperref}
\usepackage[noabbrev]{cleveref}
\usepackage{dsfont}
\usepackage{enumerate}
\usepackage[shortlabels]{enumitem}

\DeclareMathOperator{\cm}{cm}

\DeclareMathOperator{\MeV}{MeV}

\DeclareMathOperator{\s}{s}

\DeclareMathOperator{\erg}{erg}

\DeclareMathOperator{\km}{km}

\DeclareMathOperator{\g}{g}

\newcommand{\fF}{\ensuremath{f_{\text{\scriptsize F}}}}
\newcommand{\fB}{\ensuremath{f_{\text{\scriptsize B}}}}

\newcommand{\gae}{\ensuremath{\hat g_{ae}}}
\newcommand{\gaeScalar}{\ensuremath{g_{ae}}}
\newcommand{\gag}{\ensuremath{g_{a\gamma}}}
\newcommand{\gagEff}{\ensuremath{g_{a\gamma}^{\text{eff}}}}
\newcommand{\gagDec}{\ensuremath{g_{a \gamma}^{\text{(D)}}}}
\newcommand{\gagPri}{\ensuremath{g_{a\gamma}^{\text{(P)}}}}

\newcommand{\appropto}{\mathrel{\vcenter{
  \offinterlineskip\halign{\hfil$##$\cr
    \propto\cr\noalign{\kern2pt}\sim\cr\noalign{\kern-2pt}}}}}

\renewcommand\vec[1]{{\bf #1}}

\newcommand{\Neff}{N_{\text{eff}}}

\newcommand{\beq}{\begin{equation}}
\newcommand{\eeq}{\end{equation}}

\allowdisplaybreaks

\title{Strong supernovae bounds on ALPs from quantum loops}
\author[a]{Ricardo Z.~Ferreira,}
\author[b]{M.C.~David~Marsh,}
\author[b]{Eike Müller}

\emailAdd{rzambujal@ifae.es}
\emailAdd{david.marsh@fysik.su.se}
\emailAdd{eike.muller@fysik.su.se}

\affiliation[a]{Institut de F\'isica d'Altes Energies (IFAE) and Barcelona Institute of Science and Technology (BIST),
		Campus UAB, 08193 Bellaterra, Barcelona, Spain
		}
\affiliation[b]{The Oskar Klein Centre for Cosmoparticle Physics,
Department of Physics,
Stockholm University, AlbaNova, 10691 Stockholm, Sweden}

\abstract{
We show that in theories of axionlike particles (ALPs) coupled to electrons at tree-level, the one-loop effective coupling to photons is process dependent: the effective coupling relevant for decay processes, $g_{a\gamma}^{\text{(D)}}$, differs significantly from the coupling appearing in the phenomenologically important Primakoff process, $g_{a\gamma}^{\text{(P)}}$. We show that this has important implications for the physics of massive ALPs in hot and dense environments, such as supernovae. We derive, as a consequence, new limits on the ALP-electron coupling, $\gae$, from SN 1987A by accounting for all relevant production processes, including one-loop processes, and considering bounds from excess cooling as well as the absence of an associated gamma-ray burst from ALP decays.  Our limits are among the strongest to date for ALP masses in the range $0.03 \MeV < m_a< 240 \MeV$. Moreover, we also show how cosmological bounds on the ALP-photon coupling translate into new, strong limits on  $\gae$ at one loop.  
Our analysis emphasises that large hierarchies between ALP effective couplings are difficult to realise once quantum loops are taken into account.
}

\makeatletter
\gdef\@fpheader{}
\makeatother

\begin{document}

\maketitle

\section{Introduction}
    Determining the elementary particle content beyond that of the Standard Model is a central objective of contemporary fundamental physics. A powerful method is to analyse extreme astrophysical environments for signals that could call for new physics. In particular, stars collapsing into supernovae have long been recognised as an ideal probe of light, weakly interacting particles, such as axions \cite{Raffelt:1996wa}.
    
    The QCD axion is the most well-motivated proposed solution to the strong CP-problem \cite{Peccei:1977hh,Peccei:1977ur,Weinberg:1977ma,Wilczek:1977pj}, and may comprise the dark matter of the universe \cite{Preskill:1982cy,Abbott:1982af,Dine:1982ah}. Axionlike particles (ALPs) are, similarly to the QCD axion, realised as pseudo-Nambu-Goldstone bosons, and frequently appear as theoretical predictions of high-energy physics models, including string compactifications \cite{Svrcek:2006yi}, some realisations of the seesaw mechanism explaining the small neutrino masses \cite{Gelmini:1980re}, or the breaking of family symmetries \cite{Davidson:1981zd,Wilczek:1982rv}.

    ALPs can couple to all Standard Model particles through interactions that respect the shift symmetry of the theory. This leads to a rich phenomenology, which is often best explored by isolating the physical effects of each coupling in turn.
    In this paper, we derive new qualitative results and quantitative bounds for ALPs that only couple to Standard Model particles at tree-level through a derivative coupling to electrons,
    \beq
        \Delta {\cal L}_{ae} = \gae (\partial_\mu a) \bar \psi_e \gamma^\mu \gamma_5 \psi_e \, .
        \label{eq:introL}
    \eeq
    Bounds derived in this way are conservative, as the inclusion of additional couplings (most notably to photons) tend to enhance possible signals. Moreover, results derived from this low-energy effective theory apply  to UV-constructions of ``photophobic’’ ALP models that have vanishing ALP-photon couplings at tree level, see e.g.~\cite{Craig:2018kne}.
   
    ALPs interacting only with electrons through equation \eqref{eq:introL} still couple to photons at loop level. In this paper, we show that this loop effect is both theoretically subtle and has significant phenomenological consequences. Moreover, loop-induced couplings are interesting as they demonstrate how large hierarchies one can in practice achieve between the parameters in the low-energy effective theory, once quantum effects are taken into account. Indeed, loop-induced two-photon decay of ALPs coupled to electrons is of critical importance for the interpretation of experimental `direct detection' searches for ALP dark matter, as recently emphasised in \cite{Ferreira:2022egk}.
    
    In this paper, we derive a new expression for the one-loop interaction of ALPs and photons, and we show that our results have immediate implications for constraints on ALPs produced in supernovae.  The main new results of this paper are:
    \begin{itemize}
    \item \emph{Correct effective couplings.} 
    At loop level, the effective couplings are often identified from  amplitudes by matching onto an effective Lagrangian. For example,  
    the (momentum-space) amplitude for the two-photon decay of an ALP interacting with electrons through equation \eqref{eq:introL} is schematically given by,
    \beq
      \mathcal{M}_{a\gamma\gamma} =  \gagDec \, \varepsilon^{\mu\nu\alpha\beta} q_1^\alpha \, q_2^\beta  (\epsilon_1)^*_\mu (\epsilon_2)^*_\nu  \, ,
      \label{eq:Mintro}
    \eeq
    where $q_1,\, q_2$ and $\epsilon_1, \epsilon_2$ respectively denote the external photon momenta and polarisations. The coupling $\gagDec$ captures the one-loop correction originating from the electron triangle diagram. The amplitude \eqref{eq:Mintro} can be obtained from the tree-level Feynman rule of the (real-space) effective operator, 
    \beq
    \Delta {\cal L}_{a\gamma\gamma} = \frac{\gagDec}{4}\, a\, F_{\mu \nu}\tilde{F}^{\mu\nu} \, ,
    \eeq
    and the prescription should capture the physics to the one loop order. It is tempting to use this effective operator to study other processes involving photons and ALPs, e.g.~the phenomenologically important Primakoff process ($a+\psi\to \gamma+\psi$ for some ion $\psi$) that involves a vertex of an ALP with two photons (cf.~\cite{Ghosh:2020vti}). An important point of this paper is that this procedure is not correct. 
    
    The effective coupling of equation \eqref{eq:Mintro} applies to the decay process, in which the ALP and the two photons are on-shell, but is not the right coupling to use for off-shell vertices, such as the one appearing in the Primakoff process.  In \cref{sec:effectivePhotonCoupling}, we calculate the relevant diagram for off-shell particles, and show that the effective coupling for the Primakoff process, $\gagPri$, differs qualitatively and quantitatively from the decay coupling:
    $$
    \gagPri \neq \gagDec \, .
    $$
    In particular, $\gagPri$ is \emph{momentum dependent} through the Mandelstam $t$ variable, and the couplings are not simply related by a rescaling. 
    Moreover, $\gagPri$ remains non-vanishing in the limit of vanishing ALP mass, contrary to some assertions in the literature. 
    This result has important phenomenological consequences, as we demonstrate for the case of ALP production in supernovae.
    
    \item \emph{Leading supernova constraints on ALPs.}  We show in \cref{sec:ALPproduction} that even in an ALP model with no tree-level (or large-logarithmic) interactions with photons, the one-loop Primakoff and $ a \leftrightarrow \gamma \gamma $ processes can play an important phenomenological role. In fact, they lead to some of the strongest limits on the ALP-electron interaction for relatively heavy ALPs ($ m_a \gtrsim 30 $~keV).
    
    We consider two independent SN constraints in this work.
    First, relatively strongly coupled ALPs provide an additional cooling channel for the SN core, potentially shortening the neutrino burst of SN1987A contrary to observations \cite{Ellis:1987pk,Raffelt:1987yt}. This `cooling bound' is studied in \cref{sec:cooling}. In this case, the one-loop processes add to the cooling by tree-level ALP-electron interactions that recently have been studied in \cite{Lucente:2021hbp} and that we reassess here. As we will show, neither of the two contributions can be neglected, and a consistent constraint should include both.
    Second, more weakly coupled ALPs can escape the SN and decay into gamma-ray photons, some of which would have reached the gamma-ray spectrometer on the Solar Maximum Mission satellite that was taking data for 223 seconds after the initial neutrino burst reached earth. The non-observation of an excess over the background gamma-ray rate puts an additional constraint on the number of ALPs produced in SN1987A, which we study in \cref{sec:decay}. This `decay bound' has no tree-level counterpart in an ALP model with $ \gag = 0 $ in the classical Lagrangian, since in this case the decay into gamma-rays can only occur at the one-loop level.
    
    \item \emph{Technical improvements on the derivation of SN constraints on ALPs.} For the calculation of both production and re-absorption of ALPs, as well as their conversion into gamma-rays, we have made some improvements over the literature: in this work, we consider quantum statistics for all processes that happen in the SN core (especially important for processes involving electrons and positrons, which are highly degenerate), we use energy dependent quantities throughout (e.g.~for the ALP absorption rate), and we do not neglect the ALP mass. Furthermore, we use the state-of-the-art numerical SN models of \cite{Fischer:2021jfm}.
    
    \item \emph{New cosmological bounds on the ALP-electron coupling.}
    ALPs that interact with photons can be produced in the primordial plasma via inverse decays. If they decay back into photons after the start of big bang nucleosynthesis (BBN) there can be photo-dissociation of primordial elements and/or changes in the effective number of degrees of freedom at recombination. 
    In \cref{sec: Comparison with other bounds}, we use previous constraints on the ALP-photon coupling, from BBN and cosmic microwave background (CMB) data \cite{Depta:2020zbh}, to derive new bounds on $\gae$ by taking into account that it induces $\gagDec$ at the one-loop level.  
    \end{itemize}
    
    In summary, we clarify the general definition and use of the effective coupling between ALP and photon, and demonstrate how to use it in a phenomenological calculation to derive strong bounds on $ \gae $. This procedure can be adapted to take loop effects into account for all ALPs with energies $ \omega \gtrsim m_e $ in other astrophysical, cosmological, or laboratory settings.
    
\section{Effective ALP-photon interactions} \label{sec:effectivePhotonCoupling}
\begin{figure}[t]
	    \centering
	    \includegraphics[width=.4\textwidth]{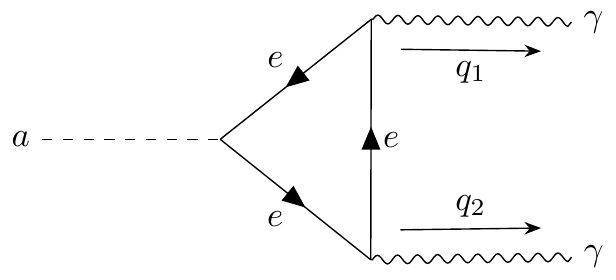}
	    \caption{Diagram contributing to the $ a\gamma\gamma $ vertex, with electrons in the loop. This diagram is finite, and therefore gives no contribution to the renormalization group equation of $ \gag(\mu) $.}
	    \label{fig:triangleDiagram}
	\end{figure}

    In this section we define and discuss the effective ALP-photon coupling generated by the electron triangle diagram shown in \cref{fig:triangleDiagram}. 
    The effective coupling depends on the four-momenta of the ALP and photons, and is sensitive to whether these states are external (on-shell) or internal (off-shell). Consequently, we show that the effective coupling for the Primakoff process differs from that of the decay process, and importantly, does not vanish for massless ALPs.  

	Throughout this paper we consider the effective theory of an ALP coupled only to electrons. For energies smaller than $\Lambda $, an appropriate UV scale, we can write the effective Lagrangian
	\begin{equation}\label{eq:Lagrangian}
    	    \mathcal{L} = -\frac{1}{2} a (\Box + m_a^2) a + \gae (\partial_\mu a) \bar \psi_e \gamma^\mu \gamma_5 \psi_e + \mathcal{L}_{\text{SM}} \, ,
	\end{equation}
	where $ a $ is a real pseudoscalar field describing the ALP of mass $ m_a $, $ \psi_e $ is the electron field, $ \gae $ is the ALP-electron coupling, and $ \mathcal{L}_{\text{SM}} $ is the standard model Lagrangian. A few comments on these conventions are in order: we define $ \gae $ with dimension $ (\text{energy})^{-1} $, which is related to the often used dimensionless coupling constant as $ \gaeScalar = 2 m_e \gae $,\footnote{Note that $$ \gaeScalar = 2 m_e \gae \simeq \frac{\gae}{1 \text{ MeV}^{-1}} $$ is just the value of the dimensionful coupling constant in units of inverse MeV. This coincidence makes numerical comparisons between references using different conventions rather simple. \label{footnote:gaeEquivalence}} where $ m_e $ is the electron mass. In most of the literature on ALPs, $ \gaeScalar $ is used because the ALP-electron interaction is often written in the non-derivative, pseudoscalar form, $ -i     \gaeScalar a \, \bar \psi_e \gamma_5 \psi_e $, which mostly leads to the same physical matrix elements at tree-level.\footnote{Even at tree-level there are counter examples, though, e.g.~if the fermion interacts with another pseudoscalar like in the case of nucleons interacting with an ALP and a pion \cite{Raffelt:1987yt,Carena:1988kr,Choi:1988xt}.}
	However, at one loop the two interaction terms are not equivalent, and the pseudoscalar interaction only matches with the derivative interaction \emph{plus} an additional direct coupling of the ALP to photons (see also \cite{Quevillon:2019zrd}).
	
	\subsection{Off-shell effective coupling}
	In the theory of equation \eqref{eq:Lagrangian}, there is no tree-level coupling to photons. However, one can still draw the loop diagram for an off-shell $ a\gamma\gamma $ vertex shown in \cref{fig:triangleDiagram} (plus the version with $ q_1 $ and $ q_2 $ exchanged), yielding the following off-shell three-point function:\footnote{As mentioned, a pseudoscalar ALP-electron coupling would lead to a different result here, even though a very similar looking one. In \cref{eq:effCouplingDefinition}, one would just replace $ 1 + 2 m_e^2 C_0 $ by $ 2 m_e^2 C_0 $, leading to the opposite behaviour of the effective coupling in the limits of infinite and vanishing electron mass.}
	\begin{equation} \label{eq:effCouplingDefinition}
	\begin{split}
	    i \mathcal{M}_{a\gamma\gamma}^{\mu\nu} &= i \frac{2 \alpha \gae}{\pi} \left[ 1 + 2 m_e^2 \, C_0\left( q_1^2, q_2^2, (q_1 + q_2)^2, m_e^2, m_e^2, m_e^2 \right) \right] q_1^\alpha \, q_2^\beta \varepsilon^{\mu\nu\alpha\beta}\\
	    &\equiv i \gagEff(q_1^2, q_2^2, (q_1 + q_2)^2) \, q_1^\alpha \, q_2^\beta \varepsilon^{\mu\nu\alpha\beta} \, ,
	\end{split}
	\end{equation}
	where $ \varepsilon $ is the Levi-Civita symbol, $ \alpha $ the fine-structure constant, and the momenta are assigned as in \cref{fig:triangleDiagram}. We used the Mathematica package FeynCalc \cite{MERTIG1991345,Shtabovenko:2016sxi,Shtabovenko:2020gxv}\footnote{We also used FeynCalc to calculate all the Feynman diagrams in the following sections, furthermore PackageX \cite{Patel:2015tea} to analytically handle loop integrals, and LoopTools to numerically evaluate them \cite{Hahn:1998yk}.} to calculate the loop diagram. We emphasise that \cref{eq:effCouplingDefinition} holds also when the particles are off-shell, and $ q_1^2, q_2^2 $ and $ q_1 \cdot q_2 $ are not fixed.  We define $ \gagEff $ as the effective ALP-photon coupling, and $ C_0 $ is the scalar, three-point Passarino-Veltman function \cite{Passarino:1978jh} for which we use the definition of \cite{Shtabovenko:2016sxi}:
	\begin{align*}
	    &C_0\left( q_1^2, q_2^2, (q_1 + q_2)^2, m_1^2, m_2^2, m_3^2 \right)\\
	    &\qquad \equiv \int \frac{\mathrm{d}^4 k}{i \pi^2} \frac{1}{\left(k^2 - m_1^2\right)\left((k - q_1)^2 - m_2^2\right)\left((k-q_1-q_2)^2 - m_3^2\right)} \, .
	\end{align*}
	The function $C_0$ can be expressed in terms of Feynman parameter integrals, and (for general arguments) as a combination of 12 dilogarithmic functions \cite{tHooft:1978jhc}. In certain limits discussed in \cref{sec:gdecay,sec:gPri}, we derive simplified analytical representation of the function. In practice, we evaluate  $ C_0 $ directly, using  the LoopTools program \cite{Hahn:1998yk}, which is sufficiently  fast for our purposes.
	
	Our expression in \cref{eq:effCouplingDefinition} agrees with similar calculations in the literature. Reference  \cite{Quevillon:2019zrd} evaluated the same diagram using Pauli-Villars regularisation, and provides a useful cross-check to our expression which is derived using the Breitenlohner-Maison-Veltman-’t Hooft scheme of dimensional regularisation \cite{Breitenlohner:1977hr,tHooft:1972tcz}, which conceptually decomposes the Dirac matrices into components in $ D $ and $ 4 $ dimensions. Furthermore, a related  discussion limited to the case of one on-shell photon with $ q^2 = 0 $ can be found in \cite{Bauer:2021mvw}, and we have explicitly checked that $ \gagEff(0, p^2, k^2) $ agrees with the Feynman parameter integral representation shown in eq.~(2.74) of that reference.
	
	The effective ALP-photon coupling of \cref{eq:effCouplingDefinition} is theoretically subtle, for several reasons:
	\begin{itemize}
        \item   First, $\gagEff$ should not be confused with the usual \emph{running coupling} $ \gag(\mu) $ that is governed by the renormalisation group (RG) equations. Running originates purely from the divergent parts of loop diagrams, yielding typically large contributions proportional to $ \log(\mu/\Lambda) $, where $ \mu $ is the renormalisation scale and $ \Lambda $ the UV cut-off of the EFT. The effective coupling includes the running coupling, but also all other loop contributions:
        \begin{equation} \label{eq:runningAndEffectiveCoupling}
            \gagEff = \underbrace{\gag(\mu)}_{\text{running}} + \underbrace{\frac{2\alpha\gae}{\pi}(1 + 2 m_e^2 C_0)}_{\text{no $\log \mu$, but $p$ dependent}} \, ,
        \end{equation}
        In the case of the ALP-photon coupling, diagram \labelcref{fig:triangleDiagram} is finite, and the running coupling $g_{a\gamma}(\mu)$ is not generated by other couplings: setting $g_{a\gamma}(\Lambda)=0$ in the UV ensures that it stays zero at lower energies \cite{Chala:2020wvs,Bauer:2020jbp}.\footnote{For a non-zero $ \gag(\Lambda) $, the dependence on the renormalisation scale $ \mu $ in \cref{eq:runningAndEffectiveCoupling} will drop out in matrix elements of physical processes if the field renormalisation constants of the external photons are included. If the loop diagram inducing $ \gag $ were divergent, the $ \mu $-dependence of the running coupling would partly be cancelled by a $ \log(\mu) $ term with the same prefactor as the $ 1/\epsilon $ pole.} However, this does not mean that there is no interaction between ALPs and photons, but only that the `leading log' contribution vanishes, and the full one-loop diagram of \cref{fig:triangleDiagram} determines the effective coupling $\gagEff$.
	
    	\item   Second, in contrast to running couplings, which only depend on the renormalization scale $ \mu $, $\gagEff$ captures the full momentum and mass dependence of the one-loop diagram. Formally, $\gagEff$ appears in the quantum effective action of the effective theory of the ALP in the operator $ \tfrac{1}{4} \, \gagEff \, a F_{\mu\nu} \tilde F^{\mu\nu} $, whose `tree-level' amplitudes reproduce the right one-loop result in \cref{fig:triangleDiagram} (cf.~e.g.~the related discussion of the 1PI effective action in \cite{Schwartz:2014sze}).  This motivates the name \emph{effective coupling} for $\gagEff$ (in a particle physics context, $ \gagEff $ could also be called a form factor). The momentum dependence of $\gagEff$ has important practical consequences for phenomenology:  
        while the running coupling is independent of the specific process in which it appears, the effective coupling is not. It is in general a function of the photons' 4-momenta $ q_1 $ and $ q_2 $ -- or more precisely the three Lorentz invariant, real quantities $ q_1^2, \, q_2^2 $ and $ (q_1 + q_2)^2 $ -- and is therefore different for each physical process involving ALPs and photons. 
	
        \item	Third,  $ \gagEff $ is not a Wilsonian effective coupling, 
        and its momentum dependence clearly does not originate from the derivative expansion of one-loop effective operators in a Wilsonian effective action. Furthermore, there are no heavy fields to be integrated out in \cref{fig:triangleDiagram} since we do not restrict ourselves to ALP masses or kinetic energies below the electron mass.
	
        \item	Fourth, $ \gagEff $ is non-vanishing even when the ALP shift symmetry is restored by taking $m_a^2 \to 0$. This should be contrasted with the discussion of \cite{Craig:2018kne}, where it was asserted that the ALP-photon coupling must scale with the order parameter of the explicit symmetry breaking, and so should vanish as the ALP mass goes to zero. We note that the symmetry of the low-energy EFT under a constant ALP shift $ a(x) \to a(x) + \delta a $ is not broken by $\gagEff\neq 0$, and the ALP-photon vertex is not forbidden on symmetry grounds. The contribution proportional to $\delta a$  is analogous to the QED $ \theta $ parameter: it is a total derivative and topologically trivial, and does not contribute to any physical quantity. Moreover, the finite $\gagEff$ is consistent with Adler's soft theorem, which states that the effective ALP-photon interaction must vanish for massless ALPs in the limit of vanishing ALP four-momentum, but does not imply a vanishing coupling at finite kinetic energy (see e.g.~\cite{Weinberg:1996kr}).

        In \cref{sec:gPri}, we show how the non-vanishing effective coupling $ \gagEff $ in the low ALP-mass limit has important implication for the Primakoff process.  
	
    \end{itemize}
	
	The loop in \cref{fig:triangleDiagram} is the vacuum or zero-temperature one-loop correction to the $ a \gamma \gamma $ three-point function. As we will discuss in \cref{subsec:ALPproduction}, one has to take finite-temperature (or chemical potential) corrections into account when considering quantum field theory in a plasma like that of SN1987A. Those effects do not change the zero-temperature loop corrections that we consider here. Especially, $ m_e $ in \cref{eq:effCouplingDefinition} is always the vacuum electron mass, even though we will discuss in \cref{subsec:bremsstrahlung} that some thermal effects can be accounted for by replacing it with an effective, thermal electron mass in tree-level diagrams. This is analogous to vacuum perturbation theory, where bare and physical mass are not equal, but it would be the bare mass appearing in the calculation of a loop.
	
	In the following two sections, we take a closer look at the effective coupling $ \gagEff $ in two specific processes relevant to our later discussion: ALP to photon decay, and the Primakoff process.
	
	\subsection{Effective coupling in ALP decays}
	\label{sec:gdecay}
    	In this section we review the calculation of the effective ALP-photon coupling relevant for the decay of ALPs into two photons.
	
	    The decay process $ a \to \gamma \gamma $ is described by the diagram in \cref{fig:triangleDiagram} with all external particles on-shell, i.e.~$ q_1^2 = q_2^2 = m_\gamma^2 $ and $ (q_1 + q_2)^2 = m_a^2 $. We include an effective photon mass $ m_\gamma $ here because we will be interested in decays inside a plasma. For these on-shell conditions, we denote the effective coupling by
	    \beq
            \gagDec \equiv \gagEff(m_\gamma^2, m_\gamma^2, m_a^2) \, .
	    \eeq
	    This `effective decay coupling' also appears in inverse decays, i.e.~photon-coalescence, where two incoming photons produce one outgoing ALP (see \cref{subfig:photonCoalescenceDiagram}).
	    
	    The effective decay coupling only depends on the masses of the ALP and the photon, and for $ m_\gamma \to 0 $ we recover the expression in \cite{Bauer:2017ris} that is used widely in the recent literature (see e.g.~\cite{Calibbi:2020jvd,Caputo:2021rux,Ghosh:2020vti,Craig:2018kne}):
	    \begin{subequations}
	    \begin{align}\label{eq:decayCoupling}
	        \gagDec &=  \frac{2 \alpha \gae}{\pi} \left[ 1 + 2 m_e^2 \, C_0\left( m_\gamma^2, m_\gamma^2, m_a^2, m_e^2, m_e^2, m_e^2 \right) \right]\\
	        \label{eq:decayCouplingNoPhotonMass}
	        &\xrightarrow{m_\gamma \to 0} \, \, \frac{2 \alpha \gae}{\pi} \left(1-\tau f(\tau)^2\right) \simeq
	        \frac{2 \alpha \gae}{\pi} \times
	        \begin{cases}
	            -\frac{1}{3} \tau^{-1} \quad &\text{for} \, \tau \gg 1\\
	            1 \quad &\text{for} \, \tau \ll 1\\
	        \end{cases} \, ,
        \end{align}
	    \end{subequations}
	    where $\tau=4m_e^2/m_a^2$, and
        \begin{equation} \label{eq:fDefinition}
        	f(\tau)= \Theta (\tau -1) \arcsin\left(1 / \sqrt{\tau} \right)+\frac{1}{2} \Theta (1-\tau ) \left[\pi +i \log \left(\frac{1 + \sqrt{1 - \tau}}{1 - \sqrt{1 - \tau}}\right)\right] \, ,
        \end{equation}
        with the Heaviside function $\Theta$.
        
        Some recent works (e.g.~\cite{Caputo:2021rux,Ghosh:2020vti}) have used $\gagDec$ to calculate other loop-level ALP-photon processes, beyond decay and coalescence. However, when the on-shell conditions are not satisfied, \cref{eq:decayCoupling} does not apply. Indeed, the correct coupling can differ significantly from $\gagDec$, as we now show for the case of the Primakoff process.
	
	\subsection{Effective coupling in the Primakoff process}
	\label{sec:gPri}
	    In this section, we calculate the relevant loop-induced coupling for the Primakoff process and explain how it differs from that of ALP decay.
	    
	    \begin{figure}
	        \centering
	        \includegraphics[width=.4\textwidth]{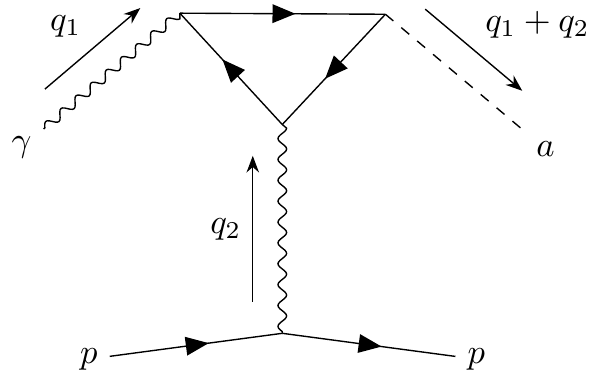}
	        \caption{Primakoff process at one loop.}
	        \label{fig:primakoffDiagramWithLoop}
	    \end{figure}
	    
	    The Primakoff process is shown in \cref{fig:primakoffDiagramWithLoop}: an on-shell photon converts into an ALP by exchanging a virtual, i.e.~off-shell, photon with a proton.\footnote{In general, all charged fermions are possible targets, but we will concentrate on protons here since that is the relevant case for the SN bounds.}
	    The relevant Lorentz invariant parameters 
	    are now given by
	    $ q_1^2 = m_\gamma^2, \, q_2^2 = t, \, (q_1 + q_2)^2 = m_a^2 $, where $ t $ is the second Mandelstam variable, and the four-momenta are as in \cref{fig:primakoffDiagramWithLoop}. We define the `effective Primakoff coupling' as 
	    \beq
	        \gagPri \equiv \gagEff(m_\gamma^2, t, m_a^2) \, .
	    \eeq
	    As long as the photon energy is small compared to the proton mass, the nuclear recoil can be neglected and the external ALP and photon energies are equal.  Under this assumption,  $ t = m_a^2 + m_\gamma^2 - 2\omega^2 (1 - \beta_a \beta_\gamma \cos\theta) $ where $\omega$ is the energy of the incoming photon (equal to the energy of the outgoing ALP), $\theta$  is the angle between the photon and ALP three-momenta, and the relativistic velocities are $ \beta_{a,\gamma}^2 = 1 - m_{a,\gamma}^2 / \omega^2 $.
	    Explicitly, the effective Primakoff coupling in the no-recoil limit is given by,
	    \begin{equation} \label{eq:effPriCoupling}
	    \begin{split}
	        \gagPri(\omega, \cos\theta) &=  \frac{2 \alpha \gae}{\pi} \left[ 1 + 2 m_e^2 \, C_0\left( m_\gamma^2, t, m_a^2, m_e^2, m_e^2, m_e^2 \right) \right]\\
	        &\xrightarrow{m_\gamma \to 0}
	        \frac{2 \alpha \gae}{\pi} \left\{1 + \frac{4 m_e^2}{m_a^2 - t} \left[f^2\left(\frac{4m_e^2}{t}\right) - f^2\left(\frac{4m_e^2}{m_a^2}\right)\right]\right\} \, ,
	    \end{split}
	    \end{equation}
	    with $ f $ as defined in \cref{eq:fDefinition}, and $ -\infty < t \leq m_a^2 + m_\gamma^2 $.
	    
\begin{figure}
	        \centering
	        \includegraphics[width=\textwidth]{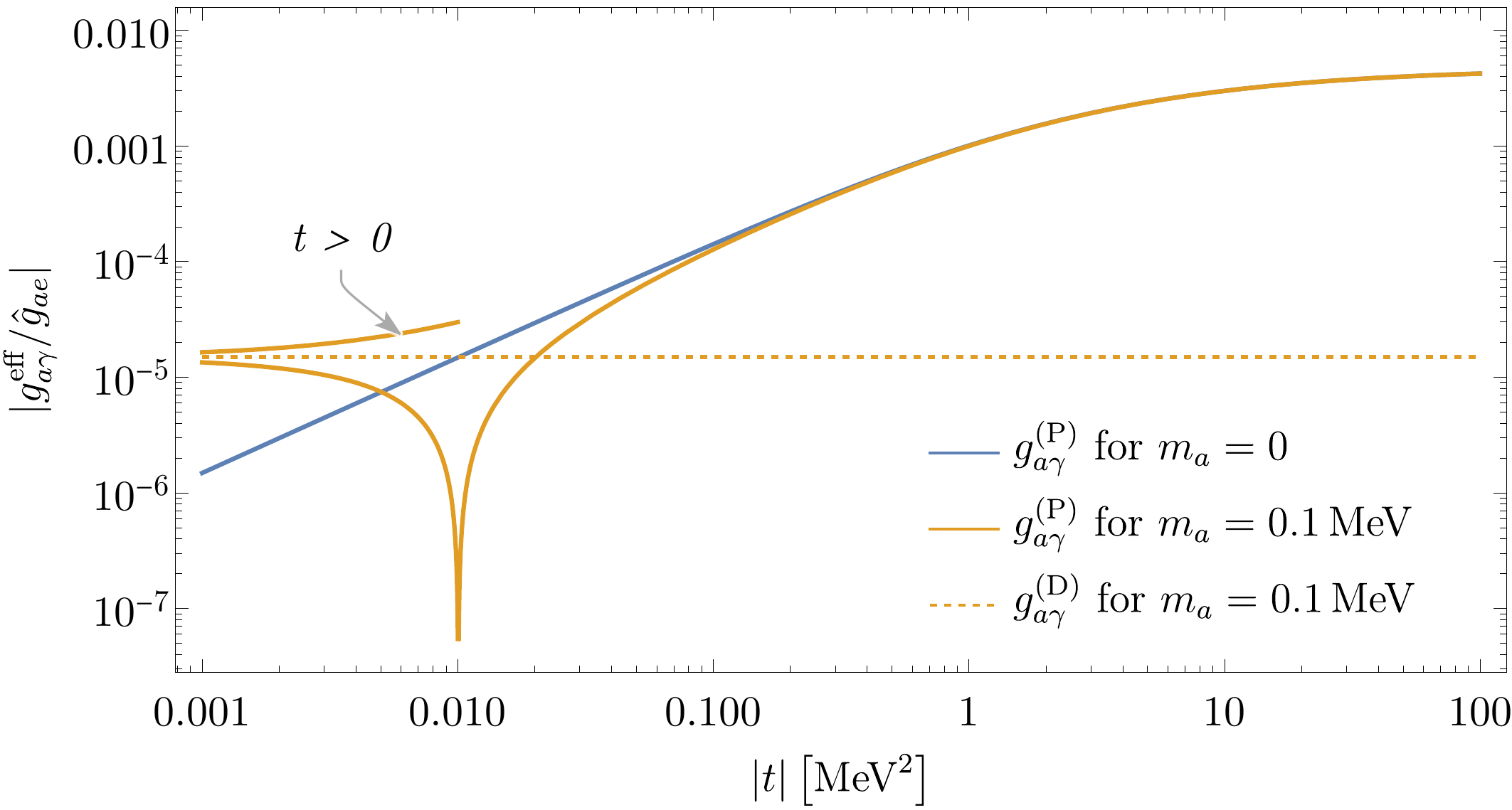}
	        \caption{Effective ALP-photon couplings for $m_\gamma=0$: $\gagPri$ evaluated
	        as function of the Mandelstam variable $ t $ for $ m_a = 0 $ (solid blue) and $ m_a = 0.1\, \text{MeV}$ (solid orange); and $ \gagDec $ for $ m_a = 0.1 \, \text{MeV}$ (dashed orange). Note that $ \gagDec = 0 $ for $ m_a = 0 $, and that $ t $ is negative unless marked differently.}
	        \label{fig:effPriCoupling}
	    \end{figure}

	    An important difference to $ \gagDec $ is that $ \gagPri $ depends on the energy transfer between photon and ALP. Therefore, while $ \gagDec \sim m_a^2 / m_e^2 $ becomes very small for low-mass ALPs, the effective Primakoff coupling remains substantial as long as the energy of the photon is larger than, or comparable to, the electron mass. \Cref{fig:effPriCoupling} shows the dependence of $\gagPri$ on $t$, and illustrates its striking difference to $\gagDec$. 
	   
	    For high-energy photons with $ \omega \gg m_e $, the effective Primakoff coupling of  \cref{fig:effPriCoupling} approaches a constant value:
	    \begin{equation}\label{eq:effCouplingSmallElectronMassLimit}
	        \gagPri \xrightarrow{\omega/m_e \to \infty} \frac{2 \alpha}{\pi} \gae \, ,
	    \end{equation}
	    This limit is relevant e.g.~when $m_a$ or $m_\gamma\gg m_e$, or when the process occurs in a highly relativistic QED plasma.\footnote{Since the evaluation of the loop function is  computationally cheap, we use the full effective coupling and do not use the constant coupling approximation for heavy ALPs or highly energetic photons when calculating ALP production rates in supernovae in \cref{sec:ALPproduction}.} Interestingly, $\gagPri$ in the limit of \cref{eq:effCouplingSmallElectronMassLimit} agrees with $ \gagDec $ in the $ m_e \to 0 $ limit. This is a consequence of the well-known anomaly: if the ALP couples to an (effectively) massless fermion, a chiral redefinition of the fermion field can eliminate the ALP-fermion coupling. However, since the PQ current is anomalous, the path integral measure is not invariant under this rotation and an ALP-photon coupling with precisely the value in \cref{eq:effCouplingSmallElectronMassLimit} is generated. Hence, in the $ m_e \to 0 $ limit, the Lagrangian in \cref{eq:Lagrangian} is equivalent to one where the ALP only couples to photons, and not the electron. Therefore, in this limit it is natural that the effective decay coupling matches with the effective Primakoff coupling.
	    
	    \Cref{eq:effCouplingSmallElectronMassLimit} also has interesting implications for more general ALP theories that include multiple charged fermions with different masses. In such theories, it is possible to assign PQ-charges to cancel the electromagnetic anomaly, in which case the total $ \gagPri $ goes to zero in the large $ t $ limit. However, even in this case, $ \gagPri $ does \emph{not} vanish at intermediate energies because the various contributions are sensitive to the ratio of $ t $ to the fermion masses.
	    
	    Due to the $ t $-dependence of $ \gagPri $, it is non-trivial to reinterpret current bounds on $ \gag $ as bounds on $ \gae $ through the loop process: most of those constraints involve integrals over a distribution of ALP-energies and the angle $ \theta $, and therefore cannot just be ``translated'' by rescaling into limits on $\gae$. Instead, the resulting processes must be re-evaluated with the $t$-dependent effective coupling. In \crefrange{sec:ALPproduction}{sec:decay}, we exemplify how this can be done when calculating ALP bounds from supernovae. 
	    
	    We close this section by briefly commenting on how our results relate to the previous literature on this subject. The loop induced coupling has been considered in  \cite{Ghosh:2020vti}, where  SN bounds on $ \gag $ were directly translated to $ \gae $ using the \emph{decay} coupling. As we have explained, this yields the wrong results for the Primakoff process: for ALP masses below $ \sim 30 $~MeV, using the decay coupling underestimates the bound by roughly a factor $ \frac{m_a^2}{12 m_e^2} $, which can be as small as $ \sim 10^{-5} $ in the mass range considered in \cite{Ghosh:2020vti}. Moreover, reference \cite{Caputo:2021rux} considered muons coupled to ALPs with either a derivative interaction proportional to $ \hat{g}_{a\mu} $ analogous to this work, or a pseudoscalar coupling $ g_{a\mu} $ as discussed below \cref{eq:Lagrangian}. For the latter, pseudoscalar case the resulting \emph{decay} coupling $ \gagDec $ was used to evaluate the one-loop Primakoff effect, and to derive constraints on $ g_{a\mu} $ from the energy loss of SNe and horizontal branch stars. While this is strictly not correct, it turns out that for a pseudoscalar muon interaction the decay coupling approximates the actual Primakoff coupling quite well, for the relevant energies and masses, and corrections are small. For the case of derivatively coupled ALPs, as they are studied here, $ \gagPri $ is not necessarily negligible, which was assumed in \cite{Caputo:2021rux}, and the inclusion of the muon-loop induced Primakoff effect would be an interesting extension of this work (see also \cref{subsec:outlook})
        
        We have shown that $\gagEff$ depends not only on the mass of the ALP, but more generally on the square of the four-momenta of ALP and photons. The effective coupling of the Primakoff process can be large if the energy of the ingoing photon is large compared to the electron mass. Thus, one should expect this one-loop effect to be particularly important in plasmas with temperatures $ T \gtrsim 1 $~MeV. An astrophysical system with such temperatures are core-collapse supernovae, and we now show how to derive new bounds from SN1987A using the loop-induced coupling.
        
\section{Production of ALPs in SN1987A} \label{sec:ALPproduction}
    Core-collapse supernovae offer an excellent opportunity to probe physics beyond the standard model because of the high temperatures and densities that are reached during the collapse. This enables exotic particles to be produced in potentially large numbers. ALPs can be produced in SNe through their couplings to any number of standard model particles like photons, electrons, nucleons, pions, or muons \cite{Ellis:1987pk,Raffelt:1987yt,Payez:2014xsa,Lucente:2020whw,Ertas:2020xcc,Lucente:2021hbp,Brinkmann:1988vi,Carenza:2019pxu,Bollig:2020xdr,Caputo:2021rux}.
    
    In this section, we first describe the state-of-the-art supernova model of \cite{Fischer:2021jfm} in \cref{subsec:SNmodel}, and provide an overview of the ALP production mechanisms in \cref{subsec:ALPproduction}. We then describe the detailed calculations for production through Bremsstrahlung in \cref{subsec:bremsstrahlung}, electron-positron fusion in \cref{subsec:eeFusionProduction}, Primakoff production in \cref{subsec:PrimakoffProduction} and photon coalescence in \cref{subsec:photonCoalescenceProduction}. The first two processes occur at tree-level, while the latter two respectively depend on the one-loop coupling $ \gagPri $ and $ \gagDec $. Our discussion on how to use the resulting spectra to constrain ALPs is deferred to \cref{sec:cooling,sec:decay}.
    
	\subsection{Supernova model}
	\label{subsec:SNmodel}
	    The production rate of ALPs in a SN plasma depends on a number of quantities, such as the temperature $ T $, the mass density $ \rho $, the electron chemical potential $ \mu_e $, the effective number of protons $ n_p^{\text{eff}} $ (a measure of number density as well as degeneracy), and the plasma frequency $ \omega_{\text{pl}} $. All these quantities vary with position and time as the SN explosion is an inhomogeneous and dynamical process. As a numerical model for those profiles, we use the reference run of the SN simulations in \cite{Fischer:2021jfm}, which uses the AGILE-BOLTZTRAN code \cite{Mezzacappa:1993gn,Liebendoerfer:2002xn}, assumes an 18 solar masses progenitor star, and spherical symmetry (i.e.~the simulations are one dimensional). The model involves six-species Boltzman neutrino transport and includes contributions from muons and their weak interactions. As in \cite{Fischer:2021jfm}, we use the relativistic mean-field nuclear equation of state \textit{DD2} as first described in \cite{Fischer:2013eka}\footnote{The equation of state data was obtained from \url{https://astro.physik.unibas.ch/en/people/matthias-hempel/equations-of-state/}, while the data of the AGILE-BOLTZTRAN simulations was kindly shared with us by Tobias Fischer.} to infer the protons' thermodynamical properties, e.g.~$ n_p^{\text{eff}} $.
	    
	    We note that the specific properties of the SN models are not tightly constrained by observations, and hence the ALP production rates can vary by a factor of up to an order of magnitude depending on the model employed, see e.g.~discussions in \cite{Ertas:2020xcc,Lucente:2020whw}. The ALP production rates scale as $ \sim \gae^2 $, and hence, the final limits on the ALP-electron coupling will have a square-root sensitivity to the uncertainties in the astrophysical production spectra.
		
		\begin{figure}[t]
		    \centering
		    \includegraphics[width=\textwidth]{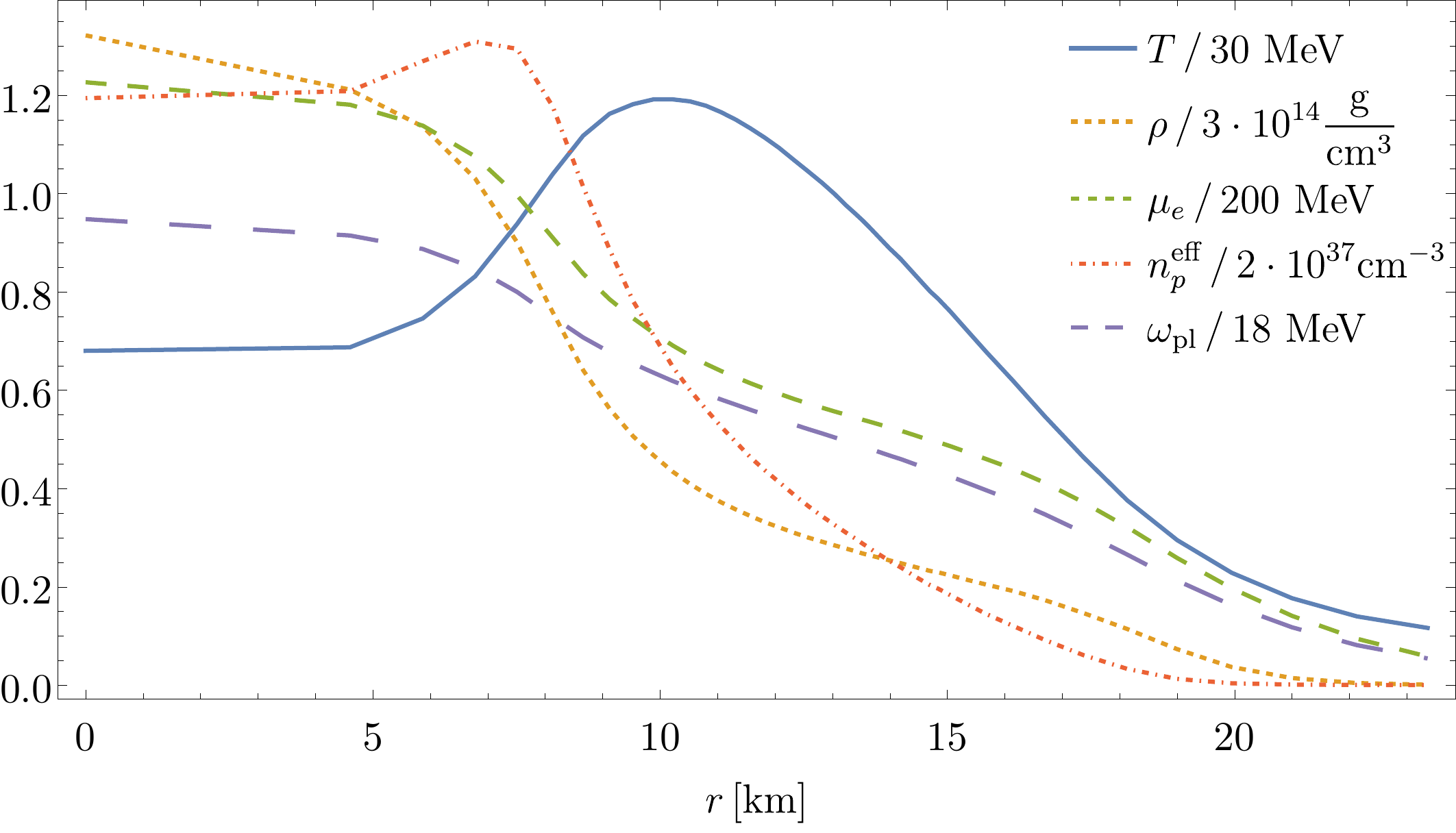}
		    \caption{Some of the SN properties at $ t = 1 $~s, taken from the simulations of \cite{Fischer:2021jfm}. At small radii, the finite resolution of the data is clearly visible.}
		    \label{fig:SNmodelPlot}
		\end{figure}
		
		\Cref{fig:SNmodelPlot} shows some normalised profiles at $ t = 1 $ second after the ``bounce'', i.e. the time at which the central density in the SN core reaches its highest value. While the matter density and electron chemical potential fall monotonically, the temperature has a maximum at $ r \simeq 10 $~km. Therefore, we will find that all ALP production processes are most efficient near this radius (at $ t = 1 $~s).
		
		In general, there will be a back-reaction of the additional cooling and energy transfer by the ALPs on the SN evolution, so that a more accurate treatment would include the ALPs and their interactions in the described SN simulations, see \cite{Betranhandy:2022bvr,Fischer:2021jfm,Fischer:2016cyd}.
		Such an analysis is, however, beyond the scope of this work, which is why we take the data of the reference run of \cite{Fischer:2021jfm}, i.e.~the simulations without any ALPs, as an SN model, and treat the ALPs as small perturbation that will not affect the relevant dynamics.
		
		With the numerical model at hand, we can now proceed to determine the flux and energy of the ALPs produced by this supernova model.

	\subsection{ALP production processes}
	\label{subsec:ALPproduction}
	    \begin{figure}
		    \centering
		    \begin{subfigure}{.38\textwidth}
		        \includegraphics[width=\textwidth]{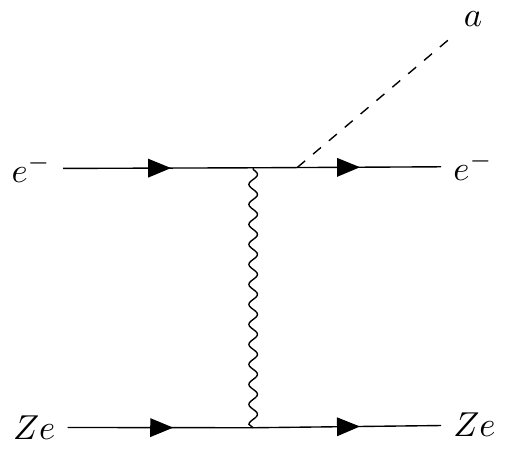}
		        \caption{Bremsstrahlung}
		        \label{subfig:bremsstrahlungDiagram}
		    \end{subfigure}
		    \hspace{.15\textwidth}
		    \begin{subfigure}{.28\textwidth}
		        \includegraphics[width=\textwidth]{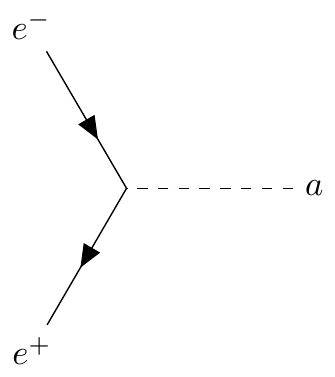}
		        \caption{Electron-positron fusion}
		        \label{subfig:electronPositronFusionDiagram}
		    \end{subfigure}
		    
		    \begin{subfigure}{.38\textwidth}
		        \includegraphics[width=\textwidth]{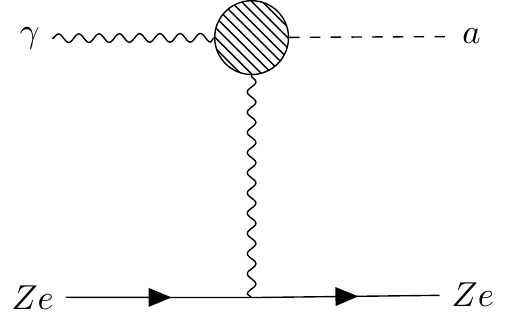}
		        \caption{Primakoff process}
		        \label{subfig:PrimakoffDiagram}
		    \end{subfigure}
		    \hspace{.15\textwidth}
		    \begin{subfigure}{.28\textwidth}
		        \includegraphics[width=\textwidth]{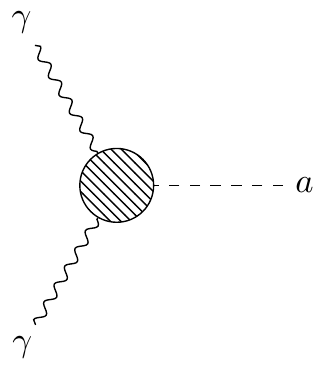}
		        \caption{Photon coalescence}
		        \label{subfig:photonCoalescenceDiagram}
		    \end{subfigure}
		    \caption{Relevant ALP production processes in SN1987A. The shaded blob in the lower two diagrams represents the one-loop, effective ALP photon interaction defined in \cref{eq:effCouplingDefinition}.}
		    \label{fig:productionDiagrams}
		\end{figure}
		
		In the following, we calculate the ALP production spectra $ \frac{\mathrm{d}^2 n}{\mathrm{d}t \, \mathrm{d}\omega} $, i.e.~the number density of ALPs produced per time and energy, of the four relevant processes in SN1987A, shown in \cref{fig:productionDiagrams}: the tree-level processes electron-nucleon Bremsstrahlung and electron-positron fusion, and one-loop processes of Primakoff production and photon coalescence. The spectra can be calculated as collision terms in the Boltzmann equation. We assume that all particles but the ALP are in thermal equilibrium, that this thermal bath is not influenced by the ALP production, and that the ALP phase space density is negligible so that no stimulated emission factor has to be included.\footnote{This is a good approximation for small couplings in the so-called free-streaming regime, but might become less suitable for larger couplings (sometimes called the trapping regime). In this part of parameter space, however, we will see in \cref{sec:cooling} that the cooling bound is controlled by the absorption rate of ALPs, and that changes of the production spectrum, e.g.~by Bose enhancement, will not affect the bound significantly.} This yields the following expression (see e.g.~\cite{Raffelt:1990yz}):
		\begin{equation}
		\begin{split} \label{eq:spectrumDefinition}
		    \frac{\mathrm{d}^2 n}{\mathrm{d}t \, \mathrm{d}\omega} =
		    \Bigg[ &\prod_i \int \frac{\mathrm{d}^3 \vec{p}_i}{(2\pi)^3 2 E_i} f_i(E_i) \Bigg]
		    \Bigg[ \prod_{j \neq a} \int \frac{\mathrm{d}^3 \vec{p}'_j}{(2\pi)^3 2 E'_j} \left[ 1 \pm f_j(E'_j) \right] \Bigg]\\
		    &\times (2\pi)^4 \delta^{(4)} \Bigg( \sum_i p_i - \sum_j p'_j \Bigg) S \,\,
		    \frac{\lvert\vec{p}'_a\rvert}{4\pi^2} \lvert \mathcal{M} \rvert^2 \, ,
		\end{split}
		\end{equation}
		where $ \vec{p}_i, \, E_i $ are the incoming particles' momenta and energies, $ \vec{p}'_j, \, E'_j $ those of the outgoing particles including the ALP, $ f_{i,j} $ are the respective phase-space distribution functions, $ |\mathcal{M}|^2 $ is the squared matrix element summed over initial and final state polarisations, and $ S = 1/n! $ is a symmetry factor to avoid overcounting of $ n $ identical particles in the initial or the final state. The quantum statistical factor for a final state particle is $ 1 + f_j $ if $ j $ represents a boson, and $ 1 - f_j $ for a fermion. Note that the product over final state momenta does not include the ALP's phase space as $ \frac{\mathrm{d}^2 n}{\mathrm{d}t \, \mathrm{d}\omega} $ is a differential with respect to the ALP's energy in the plasma frame; the difference between the differentials of energy and Lorentz-invariant phase space yields the factor $ \frac{\lvert\vec{p}'_a\rvert}{4\pi^2} $.
		
		From the production spectra, the ALP emissivity $ Q $, i.e.~the energy-loss rate per unit volume, can be determined as
		\begin{equation} \label{eq:emissivityDefinition}
                Q = \int_{m_a}^{\infty} \mathrm{d} \omega \, \omega \frac{\mathrm{d}^2 n}{\mathrm{d}t \, \mathrm{d}\omega} \, ,
		\end{equation}
		with the ALP energy $ \omega $. Note that $ Q $ in this form does not take reabsorption effects into account, see \cref{subsec:absorption} for the inclusion of those.
		
		We will use the following definitions of the Bose-Einstein and Fermi-Dirac distributions, describing the phase space densities of photons, electrons and positrons, respectively:
		\begin{equation*}
	        \fB(\omega) = \frac{1}{\exp(\omega/T) - 1} \, ,
	        \qquad
	        \fF^\mp(E) = \frac{1}{\exp[(E \mp \mu_e)/T] + 1} \, ,
		\end{equation*}
		with temperature $ T $ and electron chemical potential $ \mu_e $. We assume that electrons and positrons are in chemical equilibrium and therefore the positrons have chemical potential $ -\mu_e $.
        
        \begin{figure}
            \centering
            \begin{subfigure}{\textwidth}
                \includegraphics[width=\textwidth]{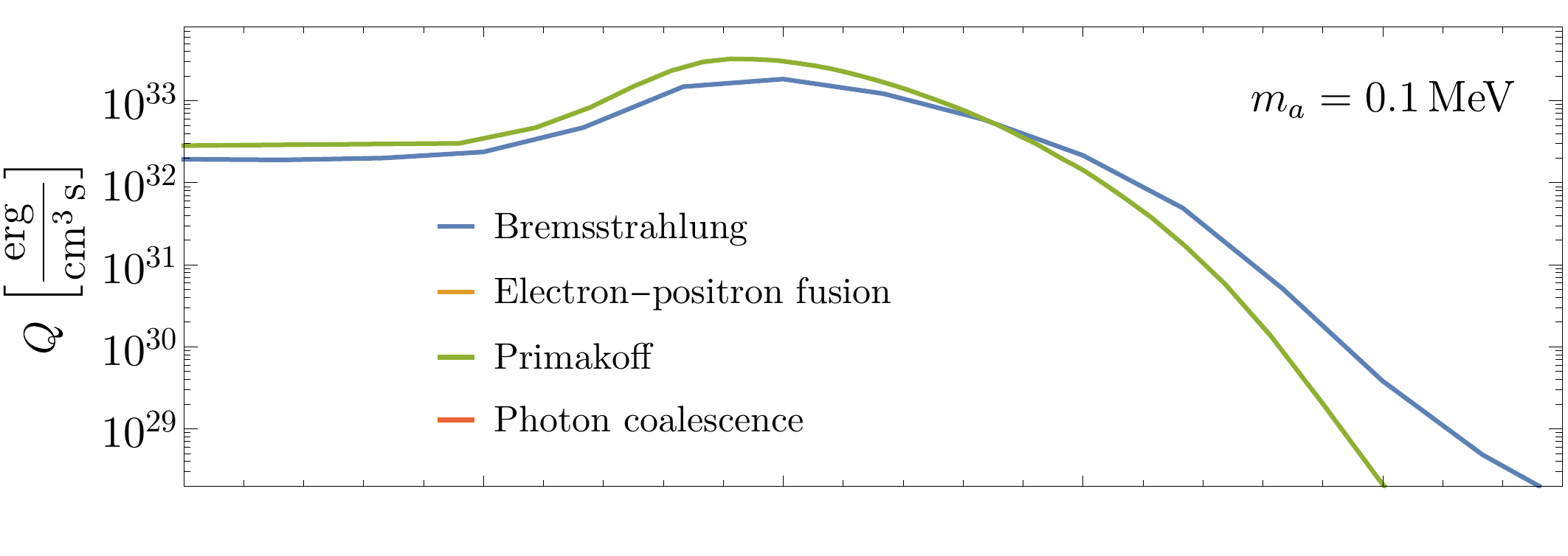}
            \end{subfigure}
            
            \vspace{-.9cm}
            
            \begin{subfigure}{\textwidth}
                \includegraphics[width=\textwidth]{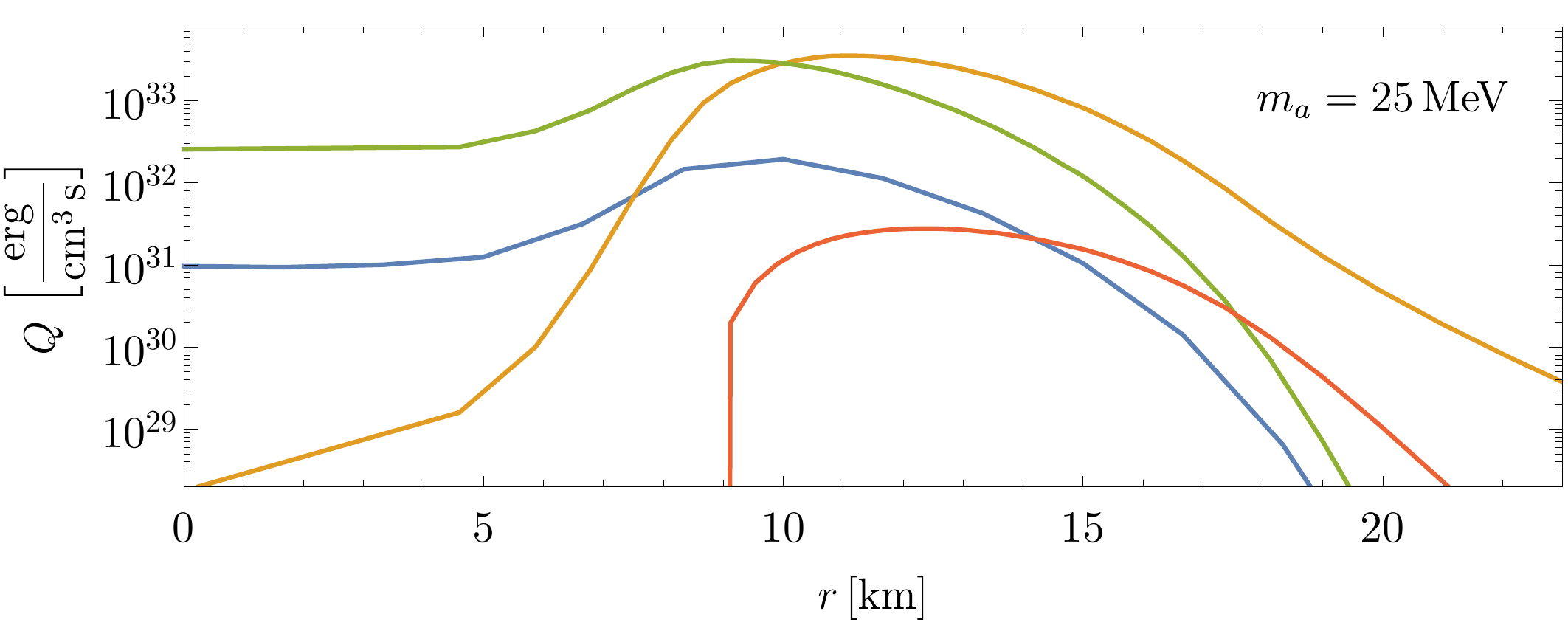}
            \end{subfigure}
            
            \caption{Comparison of ALP emissivity contributions of the four described processes for a low and a high mass, with $ \gae = 10^{-9} \text{ MeV}^{-1} $ in both plots. In the upper plot, $ m_a $ is below the threshold for both inverse decays and thus the emissivity of those processes is zero. The same happens in the lower plot for photon coalescence at $ r \leq 9 $~km.}
            \label{fig:emissivityComparisonPlot}
        \end{figure}
		
		\Cref{fig:emissivityComparisonPlot} anticipates the results of \crefrange{subsec:bremsstrahlung}{subsec:photonCoalescenceProduction}, and shows the emissivities of the four ALP production processes for different masses. One of the main, unexpected results of this paper is that the loop-induced Primakoff process results in the largest ALP emissivity and fluence at small radii (see \cref{subsec:PrimakoffProduction}). Other qualitative properties of the production spectra depend on the mass of the ALP. 	At low masses ($ m_a \lesssim 20 $~MeV), the Bremsstrahlung and Primakoff contributions dominate at all radii as  inverse decays (i.e.~electron-positron fusion and photon coalescence) are suppressed, or even kinematically forbidden. 
		At higher masses and for sufficiently large radii, the tree-level electron-positron fusion process quickly dominates as it becomes kinematically available.
		
		We now discuss, in turn, how the production spectra for Bremsstrahlung, electron-positron fusion, the Primakoff process, and photon coalescence are calculated.
        
    	\subsubsection{Bremsstrahlung}
    	\label{subsec:bremsstrahlung}
    		Electron-ion Bremsstrahlung, shown in \cref{subfig:bremsstrahlungDiagram}, is the leading production  process at tree-level for ALPs with masses $ m_{a} \lesssim 30~\text{MeV} $ in hot, degenerate plasmas \cite{Raffelt:1996wa,Lucente:2021hbp}.
    		As the upper panel of \cref{fig:emissivityComparisonPlot} shows, in our model the contribution of the tree-level Bremsstrahlung process to the ALP emissivity is dominant at large radii, i.e.~at low temperatures, but remains non-negligible at all radii.
    		
    		\Cref{eq:spectrumDefinition} yields the following expression for the production spectrum of ALPs created by electron-ion Bremsstrahlung:
    		\begin{equation}\label{eq:BremsstrahlungProduction}
    			\frac{\mathrm{d}^2 n}{\mathrm{d}t \, \mathrm{d}\omega} = \frac{n_p^{\text{eff}}}{64\pi^6} \int_{-1}^1 \mathrm{d}c_{1a} \, \int_{-1}^1 \mathrm{d}c_{12} \, \int_{0}^{2\pi} \mathrm{d}\delta \, \int_{m_{e}}^{\infty} \mathrm{d}E_2 \left\lvert \mathbf{p}_1 \right\rvert \left\lvert \mathbf{p}_2 \right\rvert\left\lvert \mathbf{p}_a \right\rvert \left\lvert \mathcal{M} \right\rvert^2 \fF^-(E_1) \left(1 - \fF^-(E_2)\right) \, ,
    		\end{equation}
    		where $ E_{1,2} $ are the incoming and outgoing electrons' energies, respectively, while the outgoing ALP has energy $ \omega $, the $ \vec{p}_i $ are the respective momenta, $ c_{1a} \, (c_{12}) $ is the cosine of the angle between the incoming electron and ALP (outgoing electron) momenta, and $ \delta $ the angle between the two planes spanned by these momentum pairs. This expression agrees with \cite{Lucente:2021hbp}. Since the ion has masses in the GeV-range, and hence does essentially not recoil when scattering with thermal electrons in a plasma with $ T \lesssim 40 $ MeV, energy conservation implies $ E_1 = \omega + E_2 $.
    		
    		Furthermore, we consider only free protons as target-ions, since electrons are highly degenerate in the SN core, and ALP production is most efficient at high temperature where no nuclear clustering occurs \cite{Lucente:2020whw}. Therefore, the number density of targets is the \emph{effective} number density of protons, taking their degeneracy into account \cite{Payez:2014xsa}
    		\begin{equation}\label{eq:npEff}
    			n_p^{\text{eff}} = 2 \int \frac{\mathrm{d}^3 p}{(2\pi)^3} f_p (1 - f_p) \, ,
    		\end{equation}
    		which can be up to $ 60\% $ smaller than the naive number density, without the factor $ (1 - f_p) $, of free protons in SN1987A \cite{Lucente:2020whw}. The effective density $ n_p^{\text{eff}} $ factors out of the expression for the spectrum in \cref{eq:BremsstrahlungProduction} because we assume no proton recoil.
    		
    		The matrix element for this process, $ \left\lvert \mathcal{M} \right\rvert^2 $, averaged over the incoming and outgoing electrons' spin, can be found in the appendix of \cite{Carenza:2021osu}, where the authors use $ k_{S}^2 T = e^2 n_p^{\text{eff}} $ and include $ n_p^{\text{eff}} $ in the definition of $ \lvert \mathcal{M} \rvert^2 $, which we do not. We used FeynCalc to reproduce and check the expression, and find agreement with \cite{Carenza:2021osu}. It should be noted that references \cite{Carenza:2021osu} and \cite{Lucente:2021hbp} used the pseudoscalar interaction between ALPs and electrons and that results are stated in terms of the dimensionless coupling $ \gaeScalar $. We checked that using the derivative coupling as included in \cref{eq:Lagrangian} leads to the same matrix element after the replacement $ \gae \to \gaeScalar / (2 m_e) $.
    		
    		The propagation of electrons and positrons in a relativistic QED plasma is significantly different from vacuum propagation. In \cite{Lucente:2021hbp} the authors argue that for ALP production in SN1987A, it is sufficient to replace the vacuum electron mass $ m_e \simeq 511 \text{ keV} $ by an effective mass depending on temperature and chemical potential:
    		\begin{equation} \label{eq:meEff}
    		    m_e^{\text{eff}} = \frac{m_e}{\sqrt{2}} + \sqrt{\frac{m_e^2}{2} + \frac{\alpha}{\pi}\left( \mu_e^2 + \pi^2 T^2 \right)} \, ,
    		\end{equation}
    		which corresponds to $ \sqrt{2} m_e^\star $ in the notation of \cite{Lucente:2021hbp}. Note that this form of the effective mass is only valid for $ m_e^{\text{eff}} \gg m_e $, i.e.~large $ T $ or $ \mu $ \cite{Braaten:1992abc}, explaining why the $ T,\mu \to 0 $ limit yields $ m_e^{\text{eff}} \to \sqrt{2} m_e > m_e $. We will use this effective mass whenever we calculate the ALP-electron tree-level processes in the SN core (i.e.~Bremsstrahlung, electron-positron fusion and their respective inverses).
    		
    		However, in \cite{Lucente:2021hbp} it was argued that the equivalence between pseudoscalar and derivative ALP-electron couplings also holds unaltered in the plasma, but remarkably, the two different couplings only lead to the same results if the replacement $ \gae \to \gaeScalar / (2 m_e) $ is done with the \emph{vacuum} electron mass $ m_e $. We note that, therefore, by using the derivative coupling and naively following the prescription of \cite{Lucente:2021hbp}, i.e.~replacing $ m_e \to m_e^{\text{eff}} $ everywhere, one would actually underestimate the Bremsstrahlung and electron-positron fusion rates by typically more than two orders of magnitude. We leave a further investigation of this puzzle to future work.
    		
    		The contribution of the Bremsstrahlung process to the ALP emissivity in SN1987A is shown in blue in \cref{fig:emissivityComparisonPlot}.
    	
    	\subsubsection{Electron-positron fusion} \label{subsec:eeFusionProduction}
    		If the ALP mass is larger than twice the effective electron mass, there is a further, and usually more efficient production process at tree-level: the inverse of the ALP-to-electron decay, electron-positron fusion. The diagram is shown in \cref{subfig:electronPositronFusionDiagram}, yielding the matrix element, averaged over initial states, $ \frac{1}{4} \lvert \mathcal{M} \rvert^2 = \gaeScalar^2 m_a^2 / 2 $. The resulting production spectrum can be calculated by integrating the electron and positron distribution functions over the positron energy $ E_{+} $
    		\begin{equation}\label{eq:electronPositronFusionSpectrum}
    	        \frac{\mathrm{d}^2 n}{\mathrm{d}t \, \mathrm{d}\omega} = \frac{\gaeScalar^2 m_a^2}{16 \pi^3} \,\Theta(m_a - 2 m_e^{\text{eff}}) \int_{E_{\text{min}}}^{E_{\text{max}}} \mathrm{d} E_+ \, \fF^-(\omega - E_+) \fF^+(E_+) \, ,
    	    \end{equation}
    	    with the ALP energy $ \omega $, energy conservation yielding $ \omega - E_{+} $ as the electron energy, and momentum conservation leading to the lower and upper integration limits
    	    \begin{equation}\label{eq:eeFusionIntegrationLimits}
    	        E_{\text{min,max}} = \frac{\omega}{2} \mp \frac{\sqrt{(\omega^2 - m_a^2)(m_a^2 - 4 (m_e^{\text{eff}})^2)}}{2 m_a} \, .
    	    \end{equation}
    	    This expression\footnote{The integral in \cref{eq:electronPositronFusionSpectrum} has an analytical solution in terms of logarithms. However, that solution yields worse results numerically due to intricate cancellations in the logarithms' argument, and we chose to evaluate the integral numerically for this work.} again agrees with \cite{Lucente:2021hbp}. Note that, as in \cite{Lucente:2021hbp}, we use the pseudoscalar coupling $ \gaeScalar $ in \cref{eq:electronPositronFusionSpectrum} following our earlier arguments (see \cref{subsec:bremsstrahlung}) about the thermal modification of the ALP electron interaction in a plasma. As also discussed there, we have to use the effective, thermal electron mass $ m_e^{\text{eff}} $ defined in \cref{eq:meEff} to take the effect of the plasma on electron and positron propagation into account.
    	    
    	    We remark here that the integrand in \cref{eq:electronPositronFusionSpectrum} peaks for small $ E_{+} $ because the positrons have chemical potential $ -\mu_e $ with $ \mu_e \gg T $ and hence their distribution function is approximately $ \fF^{+}(E_{+}) \sim \exp(-E_{+} / T) $. Therefore, in this particular case it is not clear that the modification of the positrons' dispersion relation is captured well by the simple replacement $ m_e \to m_e^{\text{eff}} $, as this approximation is valid only for hard momenta $ p_{+} = \sqrt{E_{+}^2 - \left( m_e^{\text{eff}} \right)^2} \gg m_e^{\text{eff}} $ \cite{Braaten:1992abc}.
    	    In the scope of this work, we follow the literature in using the effective mass approximation for electron-positron fusion, and leave the question of how accurate the approximation is for further study.
    	    
    	    The contribution of electron-positron fusion to the ALP emissivity in SN1987A is shown in orange in \cref{fig:emissivityComparisonPlot}. In the upper plot the ALP mass is too low for electron-positron fusion to be allowed kinematically, so the production spectrum is zero.
    		
    	\subsubsection{Primakoff process}\label{subsec:PrimakoffProduction}
    	    One of the main phenomenological result of this paper is the accurate determination of the contribution of the loop-induced Primakoff process to ALP production in hot, dense plasmas. We find that it is in fact the leading production process for $ m_a \lesssim 20 $~MeV.
    	    
    	    In the Primakoff process, a photon converts into an ALP in the field of a charged fermion, shown in the diagram in \cref{subfig:PrimakoffDiagram}. As for Bremsstrahlung, the main target fermion is a free proton, since electrons are degenerate and almost no nuclei can form in the SN core.
    	    
    		With our definition of the effective Primakoff ALP-photon coupling in \cref{eq:effPriCoupling} the differential cross-section of the process can be written compactly as \cite{Raffelt:1985nk}
    		\begin{equation}\label{eq:PrimakoffDifferentialCrossSection}
    			\frac{\mathrm{d} \sigma_{\text{P}}}{\mathrm{d} \Omega} = \left\lvert \gagPri(\omega, \cos \theta) \right\rvert^2 \frac{\alpha}{8\pi} \frac{p_\gamma p_a^3 (1 - \cos^2\theta)}{(p_\gamma^2 + p_a^2 - 2 p_\gamma p_a \cos\theta)(p_\gamma^2 + p_a^2 - 2 p_\gamma p_a \cos\theta + \kappa_{\text{S}}^2)} \, ,
    		\end{equation}
    		where we have used the limit of no ion recoil again, i.e.~$ m_{\text{ion}} \to \infty $, in which case energy conservation implies that both photon and ALP have energy $ \omega = \sqrt{p_a^2 + m_a^2} = \sqrt{p_\gamma^2 + m_\gamma^2} $, with the respective momenta $ p_{a,\gamma} $, and $ \theta $ is the angle between the momentum vectors. For processes with external photons, it is important to note that in the SN plasma on-shell photons acquire an effective mass \cite{Kopf:1997mv}: 
    		\begin{equation} \label{eq:mgEff}
    		    m_\gamma = \omega_{\text{pl}} \simeq 16.3 \MeV Y_e^{1/3} \left( \frac{\rho}{10^{14} \g \cm^{-3}} \right)^{1/3} \, ,
    		\end{equation}
    		where $ \omega_{\text{pl}} $ is the plasma frequency and $ Y_e $ the electron fraction.
    		Furthermore, \cref{eq:PrimakoffDifferentialCrossSection} includes the screening of the Coulomb-like interaction at the Debye-Hückel scale $ \kappa_{\text{S}}^2 = e^2 n_p^{\text{eff}}/T $ \cite{Raffelt:1985nk}.
    		
            The resulting ALP spectrum is
            \begin{equation}\label{eq:PrimakoffSpectrum}
                \frac{\mathrm{d}^2 n}{\mathrm{d}t \, \mathrm{d}\omega} = \frac{n_p^{\text{eff}}}{\pi^2} \frac{\omega^2 - m_\gamma^2}{\exp(\omega/T) - 1} \sigma_{\text{P}} \, ,
            \end{equation}
            where $ \sigma_{\text{P}} $ is the total cross section, and the density of target protons factors out because of the no-recoil limit as in the case of Bremsstrahlung. Note that we cannot analytically solve the integral over $ \cos\theta $ in the definition of the total cross section since the effective coupling has a non-trivial dependence on $ \cos\theta $. Therefore, we have to evaluate the implicit integral in \cref{eq:PrimakoffSpectrum} numerically.
            
            For energies $ \omega \gg m_a, m_\gamma, m_e $ we can approximate $ \gagPri \simeq \frac{2\alpha}{\pi} \gae $, and in that case give a simple expression for the emissivity
            \begin{equation}
                Q \simeq \frac{\alpha^2 \gae^2}{\pi^3} \, T^7 F\left(\frac{\kappa_{\text S}^2}{4 T^2}\right) \, ,
            \end{equation}
            where $ F $ is defined in \cite{Raffelt:1985nk}, slowly varying, and of order $ 10^{-2} $ to $ 10^{-1} $ for typical SN conditions. We see that the Primakoff emissivity depends strongly on the temperature, explaining why it can be the leading production process in the hot SN plasma, even though it is suppressed by a loop factor. In \cref{fig:emissivityComparisonPlot} the Primakoff contribution is shown in green.
            
            In the literature, one often finds a production rate $ \Gamma_{\text P} $, see e.g.~\cite{Raffelt:1996wa,Lucente:2020whw}, which is related to the production spectrum and the total cross section as
            \begin{equation}\label{eq:primakoffProdRate}
                \Gamma_{\text P} = \pi^2 \frac{\exp(\omega/T) - 1}{\beta_\gamma \omega^2} \, \frac{\mathrm{d}^2 n}{\mathrm{d}t \, \mathrm{d}\omega} = n_p^{\text{eff}} \beta_\gamma \sigma_{\text P} \, ,
            \end{equation}
            with $ \beta_\gamma $ as defined in \cref{sec:gPri}. Replacing $ \gagPri(\omega, \cos \theta) \to \gag $ in \cref{eq:primakoffProdRate}, and integrating the cross section yields the same expression for $ \Gamma_{\text P} $ as in \cite{Lucente:2020whw}.
    	
    	\subsubsection{Photon coalescence} \label{subsec:photonCoalescenceProduction}
    	    In close analogy to the electron-positron fusion process, there is also an inverse to ALP-to-photon decay called photon coalescence with the diagram shown in \cref{subfig:photonCoalescenceDiagram}. Since in a plasma we have to treat the photons as massive with mass $ m_\gamma $, the kinematics of the process are equivalent to the electron-positron fusion case with the replacement $ m_e \to m_\gamma $. However, the averaged matrix element is slightly different $ \frac{1}{4} \lvert \mathcal{M} \rvert^2 = \frac{1}{8} \left\lvert \gagDec \right\rvert^2 m_a^4 (1 - 4 m_\gamma^2 / m_a^2) $, and we find the following expression for the production rate:
    	    \begin{align} \label{eq:photonCoalescenceSpectrum}
    	        \frac{\mathrm{d}^2 n}{\mathrm{d}t \, \mathrm{d}\omega} &= \left\lvert \gagDec \right\rvert^2 \frac{m_a^4}{128 \pi^3} \left(1 - \frac{4 m_\gamma^2}{m_a^2}\right) \, \Theta(m_a - 2 m_\gamma) \int_{\omega_{\text{min}}}^{\omega_{\text{max}}} \mathrm{d} \omega_1 \, \fB(\omega_1) \fB(\omega-\omega_1)\\
    	        &\xrightarrow{\omega \gg T} \left\lvert \gagDec \right\rvert^2 \frac{m_a^4}{128 \pi^3} \sqrt{\omega^2 - m_a^2} \left(1 - \frac{4 m_\gamma^2}{m_a^2}\right)^{3/2} \exp\left(-\frac{\omega}{T}\right) \Theta(m_a - 2 m_\gamma) \, . \nonumber
    	    \end{align}
    	    The integration limits $ \omega_{\text{min,max}} $ are given by \cref{eq:eeFusionIntegrationLimits} with $ m_e \to m_\gamma $, and we show that in the limit $ \omega \gg T $, i.e.~if $ \fB $ is well approximated by a Boltzmann distribution, the result of \cite{Lucente:2020whw} is recovered. In the $ m_\gamma \to 0 $ limit this result also agrees with the expression found in the supplemental material of \cite{Caputo:2022mah}, which to our knowledge is the first time quantum statistics have been taken into account in the photon coalescence production rate. Therefore, \cref{eq:photonCoalescenceSpectrum} generalises both previous results by including an effective photon mass as well as the quantum distribution function for the photons.
    	    
    	    The ALP emissivity due to photon coalescence is shown in red in \cref{fig:emissivityComparisonPlot}. As is the case for electron-positron fusion, this inverse decay is kinematically only allowed when the ALP mass is large enough. In fact, in our SN model the effective photon mass is always larger than the effective electron mass (cf.~\cref{eq:meEff,eq:mgEff}), and hence, to be efficient, photon coalescence requires even larger $ m_a $ than electron-positron fusion. This can be seen in the lower plot in \cref{fig:emissivityComparisonPlot} with $ m_a = 25 $~MeV: the ALP emissivity due to photon coalescence goes to zero for radii below 9 km because there $ m_\gamma(r) > m_a / 2  $, while electron-positron fusion is possible for all radii.
    	    
    	    The production spectra calculated in this section will now be used to constrain the ALP-electron coupling via the so-called \emph{cooling bound} in \cref{sec:cooling}, and the \emph{decay bound} in \cref{sec:decay}.

\section{Cooling bound}
\label{sec:cooling}
    \begin{figure}
        \centering
        \includegraphics[width=.85\textwidth]{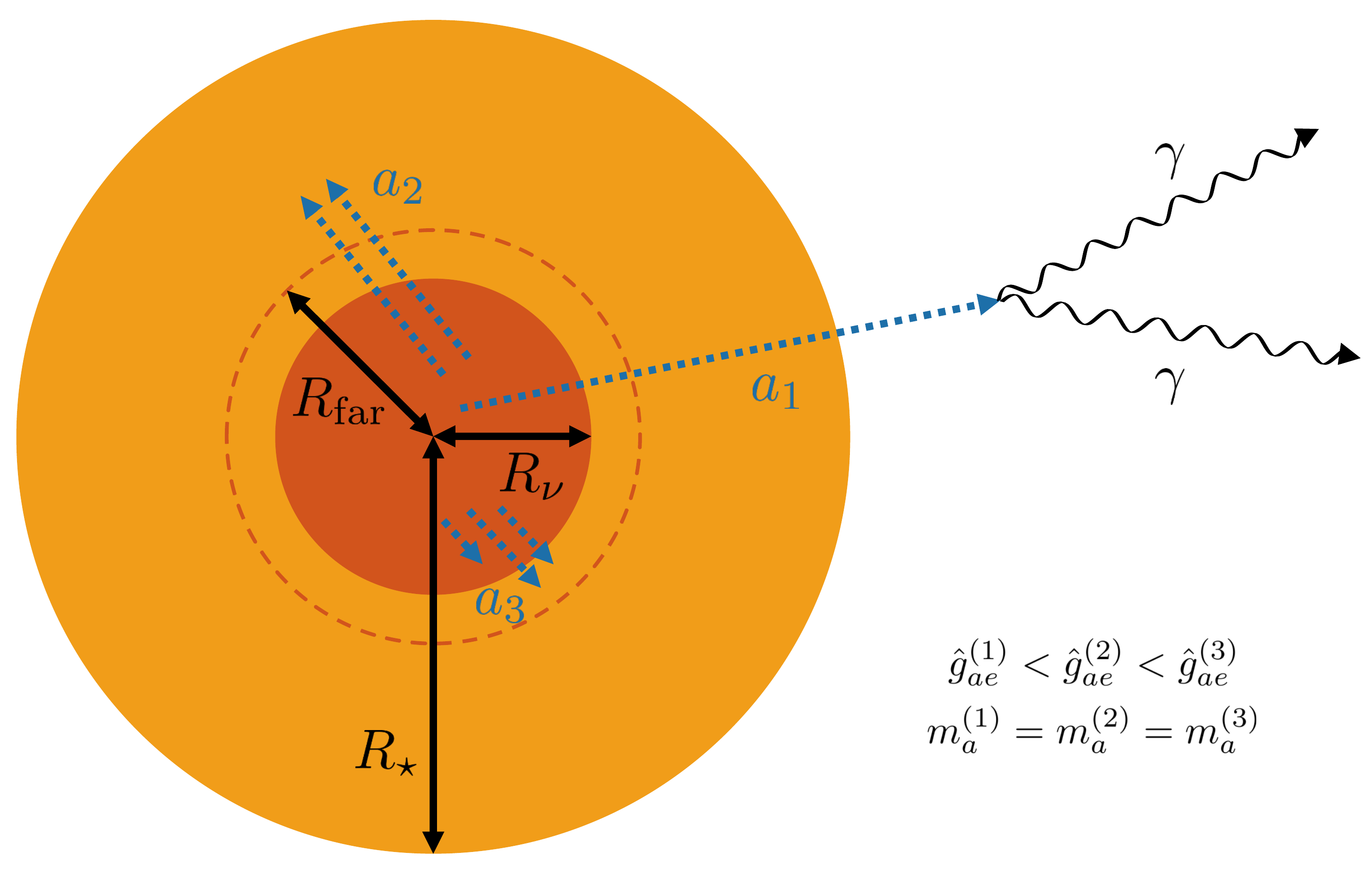}
        \caption{Mechanisms of the two supernova constraints for three different, exemplary ALP models $ a_i $ with couplings $ \gae^{(i)} $ and masses $ m_a^{(i)} $. ALPs are predominantly produced in the neutrino sphere with radius $ R_\nu $ (shown in dark orange), with the number of arrows indicating the respective flux; note that production in the actual model is isotropic and spherically symmetric. The ALPs will subsequently be reabsorbed by the SN plasma ($ a_{2} $ and $ a_{3} $) or decay into photons or electron-positron pairs where kinematically allowed (possible for all three depicted scenarios). The typical time and length scales after which reabsorption and decay happen depend on the coupling strength and, as depicted here, weaker couplings lead to longer mean-free paths. ALP $ a_1 $ will contribute both to the decay and cooling bound, $ a_2 $ only to the cooling bound since it does not escape the SN remnant with radius $ R_\star $ (shown in orange), and $ a_3 $ to neither since it is absorbed before the radius $ R_{\text{far}} $ as defined below. See full text for details.}
        \label{fig:SNFigure}
    \end{figure}
    
    An illustration of the SN constraints we consider here and in \cref{sec:decay} is shown in \cref{fig:SNFigure}. In this section, we first examine how the SN plasma loses energy by ejecting ALPs, and why too much of this cooling is incompatible with observations.
    
    The relevant observation here is that SN1987A was accompanied by a 10s long neutrino burst. As suggested by Raffelt in \cite{Raffelt:1996wa}, a simple analytical estimate shows that the duration of this burst would roughly be cut in half if there was an additional energy loss channel at $t=1\, $s after the initial bounce with a luminosity as large as the one of neutrinos $ L_\nu \simeq 3 \cdot 10^{52} \frac{\erg}{\text{s}} $. Note that $ L_\nu $ is taken from the SN simulations that we use \cite{Fischer:2021jfm}. As should be clear from the discussion in \cref{sec:ALPproduction}, weakly coupled ALPs provide such an additional energy loss mechanism if they can escape the SN. Therefore, we can derive a constraint on their interactions with the plasma, which we assume to be just due to $ \gae $, by demanding that
    \begin{equation} \label{eq:RaffeltCriterion}
    L_a < L_\nu \simeq 3 \cdot 10^{52} \frac{\erg}{\text{s}} \, .
    \end{equation}
    
    Before deriving the ALP luminosity in \cref{subsec:modifiedLuminosity}, we first consider in \cref{subsec:absorption} that they can also be reabsorbed by the plasma.
	
	\subsection{ALP reabsorption and opacity} \label{subsec:absorption}
		Due to the high temperature and large density in the supernova core, ALPs can be produced but also quickly  absorbed if the coupling to electrons or photons is large enough. If this happens in the core of the SN, they do not contribute to the energy loss that is relevant for the shortening of the neutrino burst.
		
		The relevant quantity characterising the ALP absorption is the local mean free path, defined as product of the (relativistic) velocity of the ALP $ \beta_a = \sqrt{1 - \left(\frac{m_a}{\omega}\right)^2} $ and the inverse of its reduced absorption rate $ \Gamma_{\text{abs}} $ \cite{Caputo:2021rux}:
		\begin{equation}\label{eq:mfpDefinition}
		    \lambda = \beta_a \, \Gamma_{\text{abs}}^{-1} \, .
		\end{equation}
		The reduced absorption rate is the difference between the absorption and spontaneous emission rates of the ALP, which due to detailed balance (see also \cref{eq:absorptionRateDefinition}) is equal to $ 1 - \exp(-\omega/T) $ times the absorption rate \cite{Caputo:2021rux}.\footnote{We thank Andrea Caputo for pointing out the difference between the absorption rate and its reduced version.}
		
		To calculate the total reduced absorption rate, we include the inverses of all production processes mentioned above: Primakoff, Bremsstrahlung, Photon coalescence, and electron-positron fusion, where the latter two are just decays. The rates of the respective inverse processes can be calculated from the collision term, similar to \cref{eq:spectrumDefinition}, and are therefore related to the production spectra of \cref{subsec:ALPproduction} as
		\begin{equation}\label{eq:absorptionRateDefinition}
		    \Gamma_{\text{abs}} = \frac{2\pi^2 [\exp(\omega / T) - 1]}{\beta_a \omega^2} \frac{\mathrm{d}^2 n}{\mathrm{d}t \, \mathrm{d}\omega} \, ,
		\end{equation}
		where we used the following relations to convert initial and final state distribution factors into another:
		\begin{equation*}
		    1 - \fF^\mp(E) = \exp\left(\frac{E \mp \mu}{T}\right) \fF^\mp(E) \, , \qquad
		    1 + \fB(E) = \exp\left(\frac{E}{T}\right) \fB(E) \, .
		\end{equation*}
		These hold because we assume all particles except for the ALP to be in thermal equilibrium. Note that the rate $ \Gamma_{\text{abs}} $ does not depend on the ALPs' phase-space distribution, and therefore does not receive corrections at large couplings where interactions are more efficient and would eventually bring the ALPs into equilibrium with the plasma.
		
		In the following, we determine the mean free paths of the four relevant processes and comment on some improvements over the existing calculations in the literature -- apart from the inclusion of one-loop corrections.
		
		Using \cref{eq:mfpDefinition,eq:absorptionRateDefinition,eq:BremsstrahlungProduction}, we find that inverse Bremsstrahlung has a mean free path of
		\begin{equation}
		\begin{split}
		    \lambda_{\text{B}}^{-1} &= \frac{n_p^{\text{eff}} [\exp(\omega/T) - 1]}{32 \pi^4}\\
		    &\quad \times \int_{-1}^1 \mathrm{d}c_{1a} \, \int_{-1}^1 \mathrm{d}c_{12} \, \int_{0}^{2\pi} \mathrm{d}\delta \, \int_{m_{\text{e}}}^{\infty} \mathrm{d}E_2 \frac{\left\lvert \mathbf{p}_1 \right\rvert \left\lvert \mathbf{p}_2 \right\rvert}{\left\lvert \mathbf{p}_a \right\rvert} \left\lvert \mathcal{M} \right\rvert^2 \fF^{-}(\omega + E_2) \left[1 - \fF^{-}(E_2)\right] \, .
		\end{split}
		\end{equation}
		Note that this expression differs by a factor $ \frac{1 - \exp(-\omega / T)}{\beta_a \pi} $ from the expression found in equation (31) of \cite{Lucente:2021hbp}, where the numerator is due to the fact that the reduced absorption rate has not been used in \cite{Lucente:2021hbp}.
		
		For the ALP-to-electron decay, we find
		\begin{equation}
		\begin{split}
		    \Gamma_{\text{abs}}^{a \to e^+ e^-} &= \frac{\gaeScalar^2 m_a^2 \left(e^{\omega/T} - 1\right)}{8\pi \, \omega^2 \beta_a} \,\Theta(m_a - 2 m_e^{\text{eff}})\int_{E_{\text{min}}}^{E_{\text{max}}} \mathrm{d} E_+ \, \fF^-(\omega - E_+) \fF^+(E_+)\\
		    &= \frac{\gaeScalar^2 m_a^2 \left(1 - e^{-\omega/T} \right)}{8\pi \, \omega^2 \beta_a} \,\Theta(m_a - 2 m_e^{\text{eff}})\int_{E_{\text{min}}}^{E_{\text{max}}} \mathrm{d} E_+ \, \left[1 - \fF^-(\omega - E_+)\right] \left[1 - \fF^+(E_+)\right]\\
		    &\xrightarrow{T \to 0} \gamma_a^{-1} \, \Gamma_0^{a \to e^+ e^-} \, ,
		\end{split}
		\end{equation}
		where $ \Gamma_0^{a \to e^{+}e^{-}} $ is the usual ALP decay rate in its rest frame (see e.g.~\cite{Bauer:2017ris}), and $ \gamma_a = \frac{\omega}{m_a} $ is the Lorentz factor. As explained in \cref{subsec:eeFusionProduction}, we take quantum statistics into account, which for the ALP-to-electron decay means a suppression of the rate by Pauli blocking due to the electrons present in the plasma. In the $ T \to 0 $ limit, i.e.~when using Boltzmann statistics for the electrons (and not using the reduced absorption rate), we reproduce the rate used in \cite{Lucente:2021hbp}. However, ignoring Pauli blocking in a highly degenerate plasma overestimates the decay rate severely, by up to three orders of magnitude in the inner core of the SN.
		
		In case of the inverse Primakoff effect, we find instead
		\begin{equation}
		    \lambda_{\text{P}}^{-1} = 2 \, n_p^{\text{eff}} \left(\frac{\beta_\gamma}{\beta_a}\right)^2 \, \sigma_{\text{P}}
		    = 2 \, n_p^{\text{eff}} \left(\frac{\beta_\gamma}{\beta_a}\right)^2 \sigma_{\text{P}} \, ,
		\end{equation}
		which is the same expression as in \cite{Lucente:2020whw}.
		
		For the ALP-to-photon decay, \cref{eq:absorptionRateDefinition,eq:photonCoalescenceSpectrum} yield
		\begin{equation}
		\begin{split}
		    \Gamma_{\text{abs}}^{a \to \gamma\gamma} &= \frac{\big\lvert \gagDec \big\rvert^2 m_a^2 (m_a^2 - 4 m_\gamma^2)}{64 \pi \beta_a \, \omega^2} \left(e^{\omega/T} - 1\right) \, \Theta(m_a - 2 m_\gamma) \int_{\omega_{\text{min}}}^{\omega_{\text{max}}} \mathrm{d} \omega_1 \, \fB(\omega_1) \fB(\omega-\omega_1)\\
		    &= \frac{\big\lvert \gagDec \big\rvert^2 m_a^2 (m_a^2 - 4 m_\gamma^2)}{64 \pi \beta_a \, \omega^2} \left(1 - e^{-\omega/T} \right) \Theta(m_a - 2 m_\gamma)\\
		    &\qquad \times \int_{\omega_{\text{min}}}^{\omega_{\text{max}}} \mathrm{d} \omega_1 \, \left[1 + \fB(\omega_1)\right] \left[1 + \fB(\omega-\omega_1) \right]\\
		    &\xrightarrow{T \to 0} \left\lvert \gagDec \right\rvert^2 \frac{m_a^4}{64\pi \, \omega} \left( 1 - \frac{4 m_\gamma^2}{m_a^2} \right)^{3/2} = \gamma_a^{-1} \, \Gamma_0^{a \to \gamma\gamma} \, ,
		\end{split}
		\end{equation}
		where also here $ \Gamma_0^{a \to \gamma\gamma} $ is the decay rate in the ALP rest frame (see e.g.~\cite{Bauer:2017ris}). As in the case of the electron decay, quantum statistics influence the rate, which in this case is Bose enhanced. Here, the effect is smaller than in the case of decay into electron-positron pairs, but can still enlarge the rate by a factor of 2 for typical SN values. Ignoring bose enhancement (and the reduction of the absorption rate), we reproduce the rate given in \cite{Lucente:2020whw}.
	
	\subsection{Modified luminosity criterion} \label{subsec:modifiedLuminosity}
	    As mentioned in the beginning of this section, if the ALPs produced in SN1987A carry away enough energy from the region from which the observed neutrino burst originated -- the so-called ``neutrino sphere'' with radius $ R_\nu $ -- the duration of this burst would shorten. Thus, the ALP has to couple either weakly enough so that no relevant abundance is produced in the SN, or so strongly that most of the emitted ALPs are reabsorbed by the plasma before they leave the SN core.
	    
	    To implement this bound, we follow the procedure of \cite{Chang:2016ntp}, which was adopted to ALPs in \cite{Chang:2018rso,Ertas:2020xcc,Lucente:2020whw,Lucente:2021hbp}: we integrate the energy carried away by ALPs from the neutrino sphere with radius $ R_\nu $, which are not reabsorbed before they reach the neutrino gain radius $ R_{\text{far}} $ \cite{Lucente:2020whw} (see \cref{fig:SNFigure} for an illustration of these radii). If an ALP's energy is deposited back into the plasma before it reaches $ R_{\text{far}} $, the energy could be converted into neutrinos since they are still produced efficiently in this region. Hence, the energy deposited into such an ALP does not shorten the neutrino burst and should not be included in \cref{eq:RaffeltCriterion}. The \emph{modified luminosity} $ L_a $ is the relevant quantity in this inequality; it is the cooling power of ALPs that cannot be efficiently reconverted into neutrinos.\footnote{In terms of the models illustrated in \cref{fig:SNFigure}: while scenario $ a_3 $ naively has the largest ALP emissivity as defined in \cref{eq:emissivityDefinition}, those ALPs are mostly reabsorbed in the sphere with radius $ R_{\text{far}} $ and hence do not contribute significantly to the modified luminosity. The ALPs in model $ a_2 $ on the other hand can lead to a shortening of the neutrino burst, while $ a_1 $ does escape even the progenitor star remnant, but could be too weakly coupled to shorten the neutrino burst.} The modified luminosity can be calculated as the following integral over the neutrino sphere and the ALP energy, including the absorption factor $ \exp\left[-\tau(r, \omega)\right] $ where $ \tau $ is the ALP's ``optical depth'' \cite{Chang:2016ntp}:
        \begin{equation}\label{eq:luminosityDefinition}
		    L_a = \int_0^{R_\nu} \mathrm{d} r \, 4\pi r^2 \, \ell^2(r) \int_{m_a / \ell(r)}^{\infty} \mathrm{d} \omega \, \omega \, \frac{\mathrm{d}^2 n}{\mathrm{d}t \, \mathrm{d}\omega}(r,\omega) \cdot e^{-\tau(r, \omega)} \, ,
		\end{equation}
		where we fix $ R_\nu = 21 $~km and $ R_{\text{far}} = 24 $~km following \cite{Lucente:2021hbp}. The lapse function $ \ell(r) \in [0,1] $ accounts for relativistic corrections in two ways: first, ALPs with local energy $ \omega < m_a / \ell $ will be gravitationally trapped in the SN, and second, the energy observed at infinity is red-shifted compared to the local one, while the local time is dilated \cite{Caputo:2022mah}. We take the approximated, general relativistic lapse function from the SN simulation \cite{Fischer:2021jfm}. The optical depth is an integral over the inverse local mean free path\footnote{In appendix A of \cite{Chang:2016ntp} and the adaption in \cite{Lucente:2020whw}, a pre-factor $ \left(1 - \frac{r (r-R_{\text{c}})}{2 R_\nu^2}\right) $ was added to account for trajectories of ALPs that are not exactly radially outward directed. However, as remarked in \cite{Caputo:2021rux}, the factor does not increase monotonically -- as one would expect due to the increasingly long inward trajectories an ALP can take, which should of course \emph{increase} the optical depth. Instead, the prefactor increases only up until $ r = R_c / 2 \simeq R_\nu / 4 $, and then decreases even below 1 for $ r \leq R_c \simeq R_\nu / 2 $ leading to an underestimation of the optical depth. Therefore, we neglect this factor in \cref{eq:tau}. See, however, \cite{Caputo:2022rca} for a detailed study on the effect of the non-radial trajectories that appeared while this work was being completed.}
	    \begin{equation}\label{eq:tau}
		    \tau(r, \omega) = \int_r^{R_{\text{far}}} \frac{\mathrm d \tilde r}{\lambda_{\text{total}}(\tilde r, \omega)} \, .
		\end{equation}
		We improve on the treatment of absorption in \cite{Lucente:2020whw} by integrating over the energy dependent optical depth in \cref{eq:luminosityDefinition} instead of approximating $ \tau(r, \omega) \simeq \tau(r, \langle \omega \rangle) $ in the integral, where $ \langle \omega \rangle $ is the average ALP energy (cf.~\cite{Ertas:2020xcc,Lucente:2021hbp}).
		
		\begin{figure}[hb]
		    \centering
		    \includegraphics[width=\textwidth]{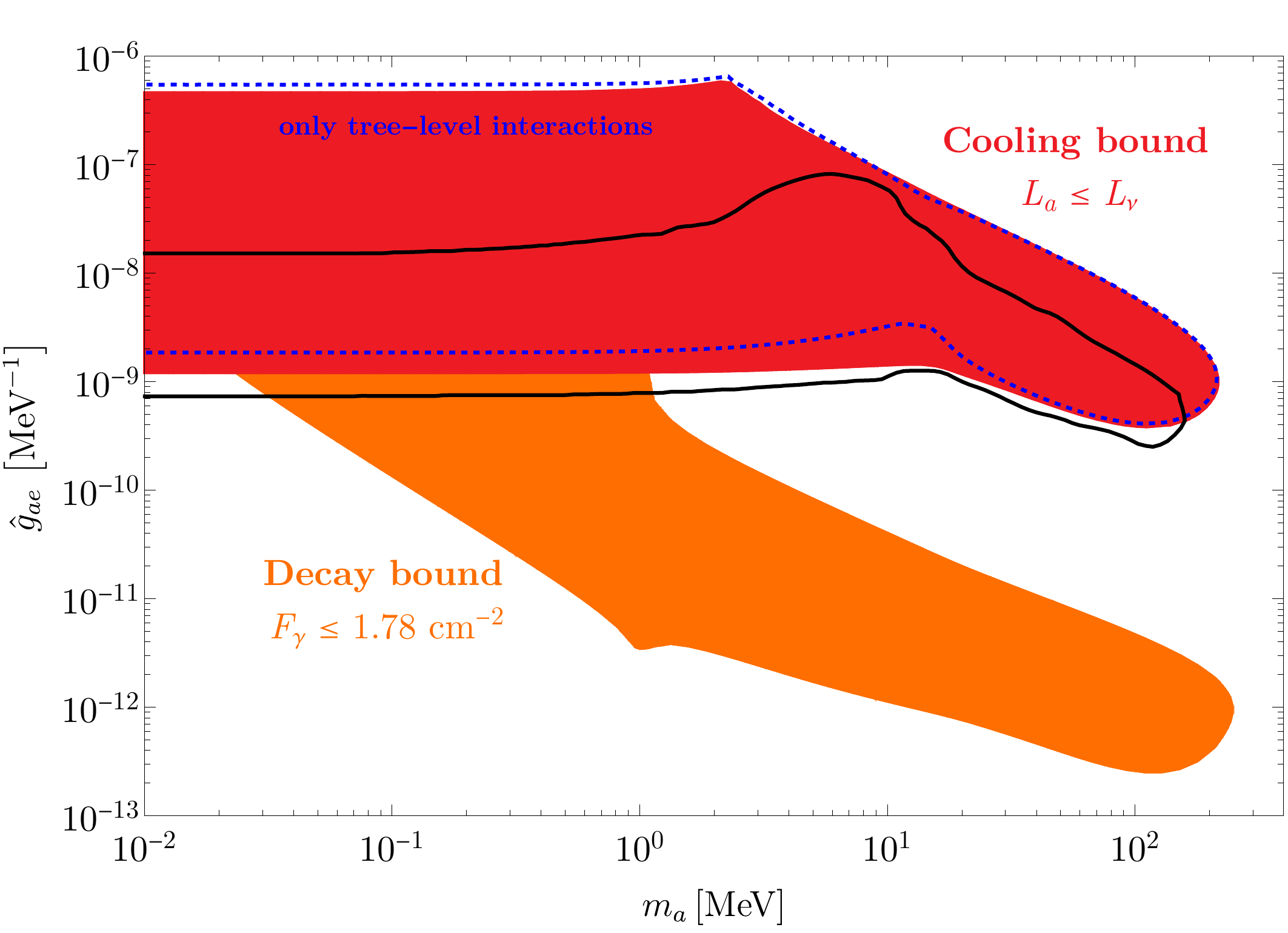}
		    \caption{Bounds on the ALP-electron coupling from anomalous cooling with only tree-level ALP-electron interactions  (dashed blue line), and including also the loop-induced ALP-photon interactions (red region), see \cref{sec:cooling}. Our bounds can be compared with the existing tree-level cooling bound from  \cite{Lucente:2021hbp} (black line).
		    Bounds from the decay into gamma-ray photons (orange region), see \cref{sec:decay}.} \label{fig:coolingAndDecayBound}
		\end{figure}
		
		We get the cooling bound on the ALP-electron coupling by demanding that the modified luminosity in ALPs should not be larger than the luminosity in neutrinos $ L_a < L_\nu \simeq 3 \times 10^{52} \text{erg} \text{ s}^{-1} $ \cite{Chang:2016ntp}. The resulting exclusion region is shown in \cref{fig:coolingAndDecayBound}.
		
		The upper boundary of this region lies in the so-called trapping regime: the ALPs are so strongly coupled that most of them are absorbed before they can escape the gain radius $ R_{\text{far}} $, decreasing the modified luminosity exponentially with increasing coupling. For large masses $ m_a \gtrsim 2 $~MeV, the dominating absorption processes are decays, mostly into electron-positron pairs, which become more efficient the heavier the ALP is. Therefore, we see that the upper end of the exclusion region decreases at larger masses. At very large masses $ m_a \gtrsim 250 $~MeV, Boltzmann suppression of the ALP production cuts off the bound as the ALPs become to heavy to be produced in a plasma with $ T \sim 30 $~MeV.
		
		Near the lower boundary of the exclusion region, the ALPs can free-stream out of the SN plasma without relevant absorption events. Here, the constraint is determined only by the production spectra of ALPs. As we have seen in \cref{sec:ALPproduction}, the production is dominated by the Primakoff process for masses of $ m_a \lesssim 20 $~MeV, while for heavier ALPs the inverse decays are more efficient.
		
		For comparison, we show the cooling bound that would result from only the tree-level ALP-electron interactions as dashed blue boundary in \cref{fig:coolingAndDecayBound}. For masses $ m_a < 20 $~MeV there is an improvement of up to a factor of 2.4 in the free-streaming regime, i.e.~at small couplings. In this part of parameter space, the Primakoff process contributes around as much to ALP production as Bremsstrahlung, see also \cref{fig:emissivityComparisonPlot}. In the trapping regime, i.e.~for large couplings, the loop-level processes seem not to contribute significantly to the modified luminosity.
		
		We also include the bound of \cite{Lucente:2021hbp} in \cref{fig:coolingAndDecayBound} in black for reference.\footnote{We thank Pierluca Carenza for helpful discussions and for providing some code of \cite{Carenza:2021osu} for easy cross-checks.} This bound is the equivalent of the blue tree-level bound calculated here. The discrepancy at large masses is due to Pauli blocking, which as we discussed in \cref{subsec:absorption} strongly suppresses the decay of ALPs to electron-positron pairs, and which was not taken into account in \cite{Lucente:2021hbp}. At small masses, the inverse Bremsstrahlung absorption process was significantly overestimated in that reference. Finally, gravitational red-shift and trapping effects were not included in the analysis of \cite{Lucente:2021hbp}, which leads to a slightly stronger bound for small couplings.

\section{Decay bound}
\label{sec:decay}
	Apart from the indirect bound on ALPs as extra cooling channel of the SN core, one can also infer a bound directly from the non-observation of gamma-rays in the energy band $ 25 - 100$~MeV after the explosion. Since the ALPs produced in the SN core via the processes studied in \cref{subsec:ALPproduction} typically have energies in this range, and can decay into photons after they leave the plasma, a certain number of gamma-ray photons should have been observed by the gamma-ray spectrometer (GRS) on board of the Solar Maximum Mission (SMM) satellite. Even though the GRS was directed towards the sun when the neutrino burst of SN1987A was observed on earth, gamma-rays from the direction of SN1987A would have hit the satellite on the side, would have penetrated the shielding, and thus still reached the detector \cite{Chupp:1989kx}. The GRS was taking data for 223 seconds after arrival of the initial neutrino burst, but went into calibration mode afterwards. One can put a bound on the number of ALPs produced in the SN since no excess gamma-rays were observed during this time \cite{Chupp:1989kx}. This bound has been considered before, most recently in \cite{Jaeckel:2017tud} where a tree-level coupling of ALPs and photons was assumed, and in \cite{Caputo:2021rux} where couplings to muons were considered at tree-level.
	
	As opposed to the cooling bound, there is no tree-level analog here for ALPs only coupled to electrons: while they can be produced in significant amounts as we have seen, at tree-level they can only decay into electrons and positrons after leaving the SN plasma. Although charged particles are not expected to reach earth, their interaction with e.g.~CMB photons or the extragalactic background light, could potentially still lead to the production of photons with $25-100$~MeV energies. These secondary processes would require a dedicated analysis. Here we will only consider the photons that result from the the decay of the ALP outside the SN via the loop process.
	
	\begin{figure}[t]
	    \centering
	    \includegraphics[width=.77\textwidth]{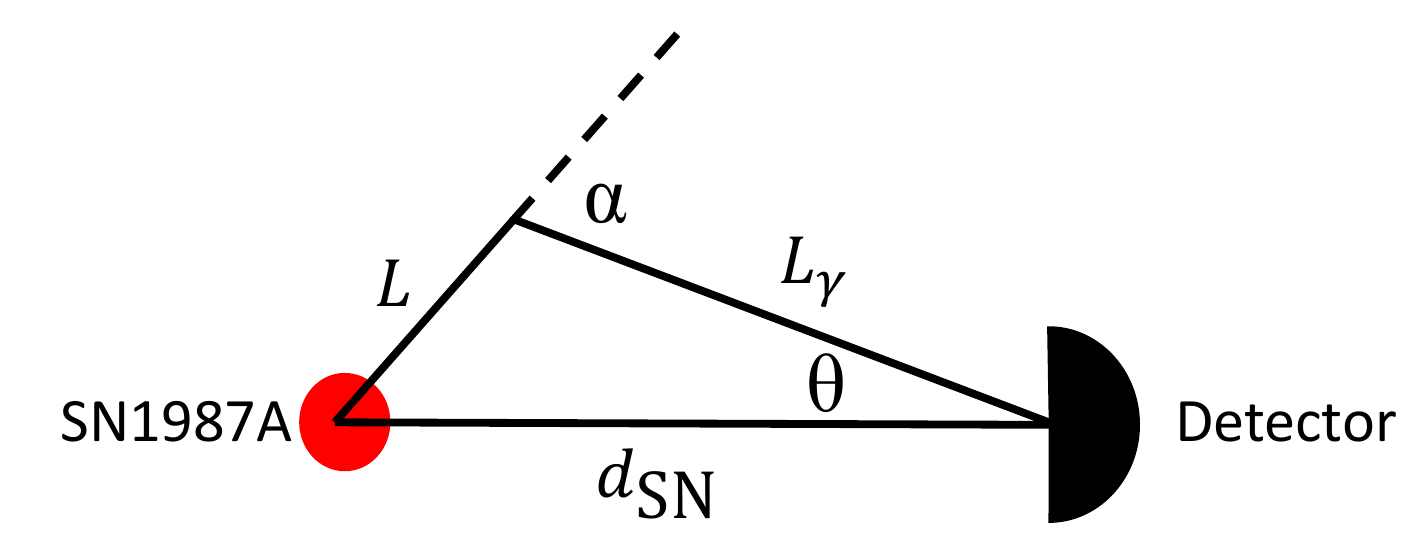}
	    \caption{Geometry of the ALP decay into a gamma-ray photon reaching earth. The ALP travels a distance $ L $ before it decays into a photon under an angle $ \alpha $. The photon travels a distance $ L_\gamma $ before it reaches the detector under an angle $ \theta $. The number of ALPs reaching the detector on the side opposite to the SN is negligible (see text) so that $ \cos \theta > 0 $, while the angles are only part of a triangle as shown if $ \cos \theta > \cos \alpha $.}
	    \label{fig:decayGeometry}
	\end{figure}
	
	The quantity we can ultimately compare to observations is $ F_\gamma $, the fluence\footnote{The fluence is the total number of photons arriving at the detector per unit of surface area.} of gamma-ray photons with energies between 25 and 100 MeV arriving at SMM in a time-window of $ \Delta t_{\text{max}} = 223s $ after the initial neutrino burst was detected. The source of $ F_\gamma $ is the ALP fluence outside of the supernova plasma. However, considering only the fluence of the source is not a good estimate for $ F_\gamma $, not even for its order of magnitude. We have to take into account several factors yielding the probability of a given ALP to decay into a photon \emph{and} for this photon to be observed by the detector. This probability distribution depends on three variables: the energy of the ALP $ \omega $, the cosine of the angle under which the ALP decays into photons $ c_\alpha \equiv \cos \alpha $, and the distance the ALP travels before it decays $ L $. Following a very similar discussion for neutrinos in \cite{Jaffe:1995sw}, we can write the differential fluence of gamma-ray photons as
    \begin{equation}\label{eq:fluenceDefinition}
        \mathrm{d} F_\gamma = 2 \cdot \mathrm{BR}_{a \to \gamma\gamma} \cdot \frac{\mathrm{d}N/\mathrm{d}\omega}{4\pi \, d_{\text{SN}}^2} \mathrm{d} \omega \cdot f_{c_\alpha}(\omega, c_\alpha) \, \mathrm{d}c_\alpha \cdot \frac{\exp[-L/l_a(\omega)]}{l_a(\omega)} \mathrm{d}L \cdot \Theta_{\text{cons.}}(\omega, c_\alpha, L) \, ,
    \end{equation}
    where the factors, in order, are
    \begin{enumerate}
        \item The number of photons produced per decay.
        
        \item The probability to decay into a pair of photons (instead of electrons), i.e.~the branching ratio. This is a constant for a given ALP mass and electron coupling.
        
        \item The spectral fluence of ALPs resulting from an isotropic production spectrum at a radius $ d_{\text{SN}} $ away from the supernova, where $ d_{\text{SN}} = 51.4 \text{ kpc} $ is the distance between supernova and earth. We can calculate this factor using the production spectra in \cref{subsec:ALPproduction}:
        \begin{equation}
            \frac{\mathrm{d} N}{\mathrm{d}\omega}(\omega) = \int_0^{R_{\text{max}}} \mathrm{d} r \, 4\pi \, r^2 \int_{t_{\text{min}}}^{t_{\text{max}}} \mathrm{d}t \, \ell^{-1}(r) \frac{\mathrm{d}^2 n}{\mathrm{d}t \, \mathrm{d}\omega_{\text{loc}}}(r,t,\ell^{-1}(r)\omega) \, ,
        \end{equation}
        where factors of $ \ell $ enter since $ \omega $ here is the energy for an observer far away (at least at a distance $ R_\star $, see below) from the SN, while the production spectra as calculated in \cref{sec:ALPproduction} are spectra with respect to the local energy $ \omega_{\text{loc}} $. The spectrum is accordingly redshifted.
        For practicality, we cut the radial integral off at $ R_{\text{max}} = 50\km $ since the contribution from high radii and therefore small temperatures and densities is negligible, and we cut the time integral off at $ t_{\text{min}} = 0.5\s $ since our SN profiles are not smooth before that time. Furthermore, we follow \cite{Jaeckel:2017tud} and approximate the production of ALPs as instantaneous by integrating the production spectrum from $ t_{\text{min}} $ to $ t_{\text{max}} = 13.3 $ s and thus neglecting the dependence of $ \mathrm{d}N/\mathrm{d}\omega $ on the delay time $ \Delta t $. Note that absorption does not play a role for the decay bound because the relevant couplings are smaller than in the cooling bound, i.e. the optical depth is negligible.
        
        \item The distribution of the angle $ \alpha $ between the ALP and photon momentum in the decay process (see \cref{fig:decayGeometry}):
        \begin{equation}
            f_{c_\alpha}(\omega,c_\alpha) = \frac{m_a^2}{2\omega^2 \left( 1 - c_\alpha \beta_a \right)^2} \, .
        \end{equation}
        Note that this is a flat distribution in the ALP's rest frame, Lorentz transformed into the frame where the ALP has energy $ \omega $.
        
        \item The probability for the ALP to decay at a distance $ L $ to $ L + \mathrm{d} L $ from the SN, where according to \cref{eq:mfpDefinition} the decay length of the ALP is
        \begin{equation}
            l_a(\omega) = \frac{\beta_a \gamma_a}{\Gamma_0^{a\to\gamma\gamma} + \Gamma_0^{a\to e^+e^-}} \, ,
        \end{equation}
        with the rest-frame decay rates $ \Gamma_0 $ (see e.g.~\cite{Bauer:2017ris}). We do not have to take into account any statistical effects like Pauli blocking or stimulated emission here, since we are interested in the case where the ALP decays after it has left the SN plasma in order for the photons to reach earth.
        
        \item The constraints we have to put on $ \omega, \, c_\alpha, \, L $, since the produced photons are detected only if a) the photon energy is in the range of the detector, b) the photon arrives on the side of the detector facing the SN,\footnote{We follow \cite{Jaeckel:2017tud} and assume that no gamma-rays arrive at the side of the detector pointing away from the SN. For low masses, this is justified for the same reason as in \cite{Jaeckel:2017tud,Oberauer:1993yr} because the decay angle distribution $ f_{c_\alpha} $ is highly peaked in the forward direction, and hence $ \theta $ in \cref{fig:decayGeometry} is small. For heavy ALPs with $ m_a > 2 m_e $ coupled to electrons, the decay channel $ a \to e^{+} e^{-} $ opens up strongly suppressing the decay length of the ALP, which for $ m_a > 1 $~MeV is always below $ d_{\text{SN}} $ in the parameter range of interest. With a short $ L $, \cref{fig:decayGeometry} again shows that $ \theta $ has to be small.} c) the photon arrives no later than 223 s after the initial neutrino burst, d) it is geometrically possible for the photon to reach earth given the values of $ m_a, \, \omega, \, c_\alpha, \, L $ (see \cref{fig:decayGeometry}), e) the ALP decays beyond the plasma surrounding the SN (so that the photons are not reabsorbed) which extends up to $ R_\star \simeq 3 \cdot 10^{7} $~km \cite{Payez:2014xsa}. This can be expressed as
        \begin{equation}
        \begin{split}
            \Theta_{\text{cons.}}(\omega, c_\alpha, L) &= \Theta(100 \MeV - \omega_\gamma(\omega, c_\alpha)) \Theta(\omega_\gamma(\omega, c_\alpha) - 25 \MeV)\\
            &\quad \times \Theta(c_\theta(c_\alpha, L)) \Theta(223\s - \Delta t(\omega, c_\alpha, L))\\
            &\quad \times \Theta(c_\theta(c_\alpha, L) - c_\alpha) \Theta(L - R_\star) \, ,
        \end{split}
        \end{equation}
        where $ \Theta $ is the Heaviside function. From the decay kinematics we get, for a given $ c_\alpha $, the photon energy
        \begin{equation}
            \omega_\gamma(\omega, c_\alpha) = \frac{m_a^2}{2\omega \left( 1 - c_\alpha \beta_a \right)} \, ,
        \end{equation}
        while from the laws of sine and cosine applied to the triangle in \cref{fig:decayGeometry} we have
        \begin{equation} \label{eq:decayGeometry}
            c_\theta(c_\alpha, L) = \frac{L_\gamma^2 + d_{\text{SN}}^2 - L^2}{2 L_\gamma d_{\text{SN}}}, \quad \text{with  } \,
            L_\gamma = L \left( \sqrt{\frac{d_{\text{SN}}^2}{L^2} - 1 + c_\alpha^2} - c_\alpha \right) \, .
        \end{equation}
        And finally, the time delay of the decay photon is
        \begin{equation}
            \Delta t (\omega, c_\alpha, L) = \frac{L}{\beta_a} + L_\gamma - d_{\text{SN}} \, ,
        \end{equation}
        taking into account that the ALP travels at a finite speed $ \beta_a < 1 $.
    \end{enumerate}
    
    For a tree-level ALP-photon coupling and including ALP masses larger than the SN temperature, the most up-to-date calculation of $ F_\gamma $ was done in \cite{Jaeckel:2017tud}. The authors used the fit of the ALP spectrum $ \frac{\mathrm{d} N}{\mathrm{d} \omega} $ of \cite{Payez:2014xsa}, which considered ultra-light ALPs, and hence the fit does not include the $ m_a $-suppression of the Primakoff effect, nor the additional production channel of photon coalescence. In \cite{Jaeckel:2017tud} the mass suppression is approximated as the ratio of the cross-sections $ \sigma(m_a) / \sigma(0) $. Here, we can improve on this by using the results of \cref{sec:ALPproduction}, i.e.~we calculate the full ALP spectrum including all relevant processes and mass dependent rates.
    
    Furthermore, the escape of ALPs from the SN and their conversion to gamma-rays is treated in \cite{Jaeckel:2017tud} by a Monte Carlo simulation, while we evaluate the analytical expression in \cref{eq:fluenceDefinition} directly. We have checked that using the ALP spectrum of \cite{Payez:2014xsa}, we can reproduce the bound of \cite{Jaeckel:2017tud} on a tree-level $ \gag $ to an accuracy of $ \mathcal{O}(10\%) $ at $ m_a \lesssim 125 $~MeV; however, at larger masses their bound seems to stop rather abruptly, while with \cref{eq:fluenceDefinition} we can extend the bound to $ m_a \lesssim 220 $~MeV.
    
    In the limit $ L \ll d_{\text{SN}} $, i.e.~when almost all decays happen close to the SN, and when additionally the ALPs are relativistic, the geometrical relations in \cref{eq:decayGeometry} simplify, the integral over $ c_\alpha $ in \cref{eq:fluenceDefinition} can be performed, and we reproduce the result of \cite{Oberauer:1993yr}, which is also used in \cite{Caputo:2021rux}. For low masses ($ m_a \ll 1 $~MeV) and a tree-level ALP-photon coupling,\footnote{The pseudoscalar ALP-muon interaction considered primarily in \cite{Caputo:2021rux} yields an essentially constant effective ALP-photon coupling for $ m_a \ll m_\mu $ at one-loop, which can therefore be treated as a tree-level coupling.} the ALPs are relativistic and $ L \ll d_{\text{SN}} $ is indeed a good approximation. In our case, however, $ \gagDec \sim m_a^2 / m_e^2 $ at low masses and thus the decay length is usually longer than the distance between SN and earth. Therefore, we cannot follow the simplifications done in \cite{Oberauer:1993yr,Caputo:2021rux}, even for low ALP masses.
    
    For $ m_a > 2 m_e $, the $ a \to e^{-} e^{+} $ decay channel opens up, and the decay length suddenly decreases below $ d_{\text{SN}} $, as long as $ \gae \gtrsim 10^{-15} $~MeV. Note that above the decay threshold the branching ratio for decays into photons suddenly decreases; the combined effects counteract each other and the resulting gamma-ray fluence is relatively continuous around $ m_a \simeq 1 $~MeV.

    Following \cite{Jaeckel:2017tud}, we use the statistical 3-$\sigma$ limit on background fluctuations of the gamma-ray fluence, $ F_\gamma < 1.78 \, \cm^{-2} $,  as an upper bound on the ALP-induced fluence. Evaluating \cref{eq:fluenceDefinition} numerically, we obtain the `decay bound' on $ \gae $ shown in orange in \cref{fig:coolingAndDecayBound}.
			
\section{Comparison with other bounds on \texorpdfstring{$\gae$}{g\_ae}}
\label{sec: Comparison with other bounds}

Cosmological data (from the CMB and BBN) and laboratory experiments can probe $ \gae $ in the mass range $m_a \gtrsim 0.01$~MeV, and provide constraints that are complementary to those from SN1987A. In \cite{Ferreira:2022egk}, we showed that the loop-induced ALP-photon coupling $\gagDec$ has significant impact on the viability of direct detection searches for ALP dark matter coupled to electrons. In this section, we conservatively assume a vanishing initial, cosmic abundance of ALPs at temperatures around 10 GeV \cite{Depta:2020zbh}. We derive new limits and summarise existing bounds on ALPs that couple to electrons.

\begin{figure}
	\centering
	\includegraphics[width=\linewidth]{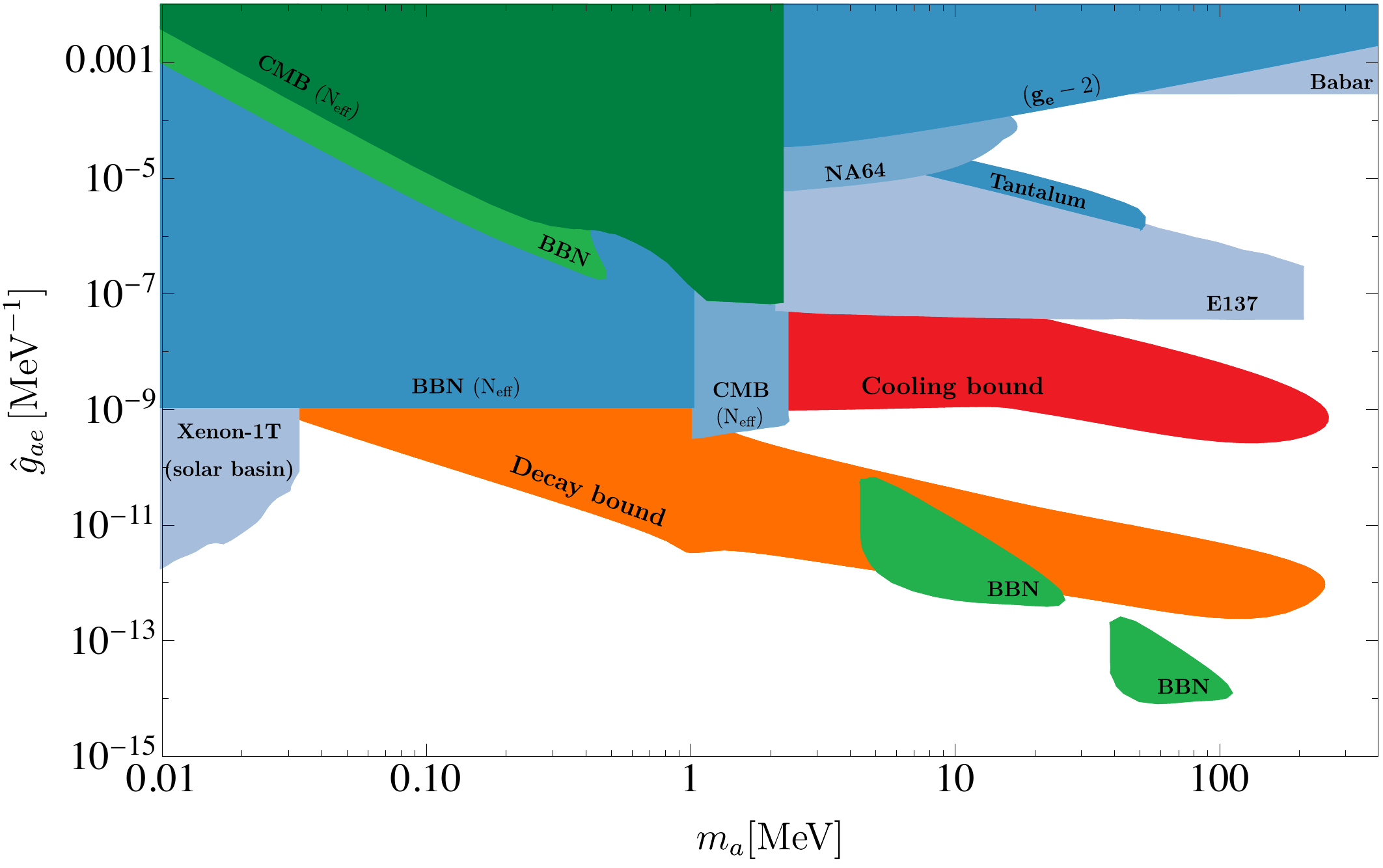}
	\caption{Compilation of laboratory, astrophysical and cosmological bounds on $\gae$. The bounds derived in this work are the SN bounds in orange and red, and the cosmological bounds in green. The remaining bounds (blue) were taken from \cite{Ghosh:2020vti,VanTilburg:2020jvl,Aguilar-Arevalo:2021wjq,Darme:2020sjf,NA64:2021aiq,Morel:2020dww,Essig:2010gu,Depta:2020zbh} (see main text for more details). Note that $ \gae $ in units of $ \MeV^{-1} $ is approximately equal to $ \gaeScalar $, see \cref{footnote:gaeEquivalence} on \cpageref{footnote:gaeEquivalence}.}
	\label{fig:Constraints on gae}
\end{figure}

We first give an overview of the most competitive laboratory and cosmological bounds on $\gae$. Then, from the current bounds on $\gagDec$, we derive new constraints on $ \gae $ using the relation between the two couplings in \cref{eq:decayCoupling}. A summary plot with all the constraints is given in \cref{fig:Constraints on gae}. New constraints on the $\gae-m_a$ parameter space derived in this work are shown in the orange and red for the supernovae constraints, and in shades of green for our cosmological constraints. Limits already present in the literature are shown in blue shades. We finish this section with a comparison between the different bounds and a discussion of their model independence.

\subsection*{Laboratory bounds}

The most competitive laboratory bounds on $\gae$ for ALP masses between $0.01-300$~MeV are:
\begin{itemize}
\item
\textit{Electron anomalous magnetic moment:}
A massive ALP coupled to electrons contributes to the anomalous magnetic moment $a_e=(g_e-2)/2$ of the electron at one-loop level as $\Delta a_e \approx \gae m_e^4/(2m_a^2 \pi^2) \left(2 \log(m_a/m_e)-11/6 \right)$ for $m_a \gg m_e$ \cite{Bauer:2021mvw}. The experimental bound $\Delta a_e<9.8\cdot 10^{-13}$ reported by \cite{Morel:2020dww} can then be translated into a bound on $\gae$.

\item
\textit{Xenon1T:}
The underground Xenon1T detector is sensitive to the local density of weakly interacting particles coupled to electrons \cite{XENON:2019gfn,XENON:2020rca}. 
Massive ALPs produced in the sun via the $\gae$ coupling can accumulate in gravitationally bound orbits around the sun and yield a density that can be constrained with Xenon1T \cite{VanTilburg:2020jvl}. 

\item
\textit{Tantalum target:}
The authors of \cite{Bechis:1979kp} reported the data analysis of a $45$ MeV electron beam hitting a tantalum target. ALPs produced by electron Bremsstrahlung could decay into electrons and positrons and be detected by the Sodium crystals. We report the bound on $\gae$ obtained in \cite{Aguilar-Arevalo:2021wjq}. 

\item
\textit{Babar:}
Searches from the Babar experiment ($e^+e^-$ collider) at the  PEP-II B-factory \cite{BaBar:2017tiz} for single photon events with missing energy. We report the constraint on $\gae$ from the analysis in \cite{Darme:2020sjf}.

\item
\textit{Beam Dumps}: 
Searches for neutral particles produced by electron/proton beams that decayed into electron-positron pairs or led to missing energy:
	
	\begin{itemize}[noitemsep,topsep=0pt,leftmargin=15pt]		
    \item{\textit{NA64}:} 
    100 GeV electron beam at CERN SPS H4 \cite{NA64:2021aiq}.
     
    \item{\textit{E137}: 20 GeV electron beam at SLAC  \cite{Bjorken:1988as,Essig:2010gu}}. 
    \end{itemize}
\end{itemize}

\subsection*{Cosmological bounds}
We now turn to bounds on $\gae$ obtained from cosmological data. ALPs will inevitably be produced in the primordial plasma via their coupling to electrons or the (loop-induced) coupling to photons. Therefore, even if their initial abundance, here considered to be at temperatures around $10$~GeV \cite{Depta:2020zbh}, is negligible, they can affect cosmological observables in several ways.
We will separate the discussion in two parts depending on whether the ALPs are produced in the primordial plasma mostly via scatterings with electrons or via inverse decays of electron-positron pairs or photons.

\vspace{0.2cm}
\textit{Production via scatterings:}
ALPs described by \cref{eq:Lagrangian} can be produced in the primordial plasma via electron-positron ($e^-e^+\to a\, \gamma$) and Compton ($e^\pm \gamma\to a\,e^\pm$) scatterings at a rate proportional to $\gae^2$. At the time of big bang nucleosynthesis (BBN), the resulting ALP abundance will contribute to the effective number of relativistic degrees of freedom, $\Neff$, that is well constrained during that epoch to be $\Neff^\text{BBN}<3.43$ at 95\% C.L. \cite{Fields:2019pfx}.

In order to derive a constraint on $\gae$ from the bound on $\Neff$, we computed the scattering cross-sections for massive ALPs (see \cref{app: Cross-sections}). We then performed the thermal average and solved the Boltzmann equation for the ALP abundance following the steps described in \cite{Ferreira:2018vjj,DEramo:2018vss}. We stopped the Boltzmann equation at neutrino decoupling ($T \simeq 2$ MeV) and translated the ALP abundance into $N_{\text{eff}}^{\text{BBN}}$ (see \cite{Ferreira:2018vjj,DEramo:2018vss} for more details). Imposing the observational bound on $N_{\text{eff}}^{\text{BBN}}$ yields the blue region labelled BBN $(\Neff)$ in \cref{fig:Constraints on gae}.
Our results are in good agreement with those obtained in \cite{Ghosh:2020vti}. Note that we restricted ourselves to the range $m_a<2m_e$: for heavier ALPs the production via inverse decays is more efficient and the discussion in the next paragraph holds. 

\vspace{0.2cm}
\textit{Production via inverse decays:}
ALPs that decay to Standard Model particles after BBN can: \emph{i)} photodisintegrate light nuclei and affect their primordial abundances \cite{Kawasaki:2017bqm,Depta:2020zbh}; \emph{ii)} heat the plasma after neutrinos have decoupled thus effectively decreasing $N_{\text{eff}}$ at the time of CMB formation \cite{Depta:2020zbh}.

A recent analysis constrained the decay rate of ALPs to electron-positron pairs, using the Planck constraint on $\Neff$, and to photons, using the Planck constraint on $\Neff$ and the measurements of primordial abundances \cite{Depta:2020zbh}. The former can be directly translated into constraints on $\gae$ (dark green region labelled CMB $(\Neff)$ in \cref{fig:Constraints on gae}). For the latter, even if the ALPs do not couple at tree-level to photons, they can decay via an electron-loop with the effective coupling $ \gagDec \sim \gae $ as in \cref{eq:decayCoupling}. Therefore, the Planck constraints on ALP-to-photon decays derived in \cite{Depta:2020zbh} also put new bounds on $ \gae $.\footnote{Note that we have only considered the loop-induced processes: ALP production via photon coalescence and ALP decay to two photons. 
For $m_a>2m_e$, the ALP production via electron-positron fusion will further increase the ALP abundance at BBN, but the possibility for the ALPs to decay into electron-positron pairs will also suppress the branching ratio for ALP decays into photons after BBN. A dedicated analysis of the combined effect is however beyond the scope of this work.}.
The excluded regions are shown in \cref{fig:Constraints on gae} in light and dark green, labelled BBN and CMB $(\Neff)$ depending, respectively, on whether the constraints on $\gagDec$ come from measurements of the primordial abundances of light nuclei or from the Planck constraint on $\Neff$. 
For larger masses and smaller couplings, the production of ALPs is less efficient so there is a smaller ALP abundance after BBN. Nevertheless, for ALP masses above twice the threshold to disintegrate deuterium ($m_a>4.4$~MeV) and Helium-4 ($m_a>39.6$~MeV), the photons resulting from ALP decays can still significantly affect the abundance of deuterium thus giving rise to the two green exclusion islands in \cref{fig:Constraints on gae}.
Note that if an initial ALP abundance is assumed stronger bounds can be derived \cite{Cadamuro:2011fd,Depta:2020zbh}.

\subsection*{Discussion of the bounds obtained}
\Cref{fig:Constraints on gae} summarises the three different types of constraints on $\gae$ that we have discussed in this paper: the SN bounds derived in \cref{sec:cooling,sec:decay} (in red and orange); the cosmological bounds derived in this work and described in the previous paragraphs (in the two shades of green); and the laboratory and cosmological bounds that were already present in the literature and also listed in the beginning of this section (in blue shades).
We will now compare the different bounds, and then discuss the model dependence of the SN bounds. 

The analysis in \cite{VanTilburg:2020jvl}, that used the data collected by the Xenon1T collaboration \cite{XENON:2019gfn,XENON:2020rca}, provides the strongest constraint on $\gae$ for ALP masses not far from the solar temperature ($m_a<30$~keV). The remaining laboratory bounds \cite{Aguilar-Arevalo:2021wjq,Darme:2020sjf,Essig:2010gu} provide competitive constraints on $\gae$ for $m_a>2m_e$ and exclude, in some regions, couplings larger than $\gae \gtrsim 4 \times 10^{-8}$.
The region constrained by ALP production from electron scatterings (region labelled BBN ($\Neff$), see also \cite{Ghosh:2020vti}) is competitive with the SN cooling bound for $m_a<2m_e$, however, it extends to much larger couplings.
At those masses, larger couplings are further excluded by the possibility of decay to photons via \cref{eq:decayCoupling} (regions labelled CMB ($\Neff$) and BBN). The region excluded by BBN also includes two islands at masses around $m_a \simeq 4-25$ and $m_a \simeq 40-110$ MeV and smaller values of the couplings, down to $\gae\simeq 10^{-14}$. Finally, the SN cooling and decay bounds derived in \cref{sec:cooling,sec:decay} constrain unique regions of parameter space,  $m_a \simeq 2-250$ MeV and $m_a \simeq 0.03-240$ MeV, respectively, that span roughly four (two) orders of magnitude in the coupling in the case of the decay (cooling) bound.

\vspace{0.2cm}
In our analysis, we isolated the physical effects originating from the ALP-electron coupling and so neglected other possible signals that do not depend on $\gae$. However, in many ALP models the presence of couplings to other standard model fields or to a dark sector is expected. Therefore, it is instructive to discuss the model dependence of the SN bounds on $\gae$ derived in \cref{sec:cooling,sec:decay} (c.f.~\cref{fig:coolingAndDecayBound}).  

Additional couplings of the ALP to particles in the core of the SN (e.g.~tree-level couplings to photons or nucleons) can have two different effects. First, they will make the production of ALPs more efficient. Therefore, the lower limit of the cooling bound derived in this work is conservative and generically model independent. Second, they will also make it harder for the ALPs that are produced to escape the SN. This will tend to decrease the upper limit of the cooling bound, the so-called trapping region. Conversely, the upper limit of the cooling bound will tend to increase if there are couplings that allow the luminosity stored in the ALPs to be converted into a dark sector before the ALP gets re-absorbed in the SN core.

For the decay bound, additional couplings to particles heavier than the ALP (e.g.~nucleons) will only improve the constraint since more ALPs can be produced, but the decay length and the $ a \to \gamma \gamma $ branching ratio in \cref{sec:decay} will not change. If, on the other hand, these particles are so light that the decay length of the ALP or the branching ratio of photon-decays will decrease, the upper end of the decay bound will be moved down to smaller couplings.

\section{Conclusion}
	 
In this section, we conclude by summarising the main results obtained in this work, and by indicating a few directions for future research.
 
\subsection{Summary of results}
	 
	The most important result of this work is that the effective ALP-photon coupling at one loop, $\gagEff$, is process-dependent (cf.~\cref{eq:effCouplingDefinition}). In the previous literature, the on-shell limit of this effective coupling, $\gagDec$, has been broadly used. While $\gagDec$ is appropriate for e.g.~the radiative decay of ALPs into photons, it is not the correct coupling for processes involving one or more off-shell particles. For example,  the Primakoff process involves one internal, off-shell photon and leads to an effective coupling, $\gagPri$, that differs qualitatively and quantitatively from $\gagDec$. While $\gagDec$ goes to zero as $m_a\to 0$, $\gagPri$ remains finite. While $\gagDec$ is momentum independent, $\gagPri$ depends on the Mandelstam $t$ variable. The difference between the couplings becomes particularly pronounced  in hot and dense environments, such as that of a supernova.
	
	Using a state-of-the-art supernova model \cite{Fischer:2021jfm}, we have shown that  quantum loops have strong implications for the predictions of ALP theories. 
	There are new channels for  ALP production at one loop (Primakoff and photon coalescence), and we showed that, due to the large temperature in the supernova core, these channels can be as efficient as the previously studied tree-level channels for ALP production. In a region of the parameter space, the ALPs can escape the supernova and carry  substantial amounts of energy away from it. This excess cooling can be constrained using the duration of the neutrino burst from SN1987A \cite{Raffelt:1996wa}. We use this method to place a `cooling bound' on $\gae$, that incorporates both tree-level and loop-level interactions. Finally, we derive a `decay bound' on $\gae$ by including the one-loop decay into photons, which can be constrained by the absence of a gamma-ray burst associated with SN1987A \cite{Jaeckel:2017tud}.
	
    Moreover, we have shown that one-loop processes lead to strong limits on  $\gae$ by considering cosmological limits from the CMB and BBN, in particular on the primordial abundance of light elements. In sum, the constraints derived in this work are the leading constraints on the ALP-electron coupling  in most of the mass range $ 0.03 \text{ MeV} < m_a < 240 \text{ MeV} $.

    Finally, we have also corrected and improved some technical aspects in the literature, such as including quantum statistics in the absorption processes in \cref{subsec:absorption}, which is important in the hot and degenerate supernova core, or improving the analysis of the probability of photon detection from ALP decays in \cref{sec:decay}.
    
	\subsection{Outlook} \label{subsec:outlook}
	There are several possible interesting extensions of this work.
	 
	With the formulas for the differential fluence of gamma-rays in \cref{eq:fluenceDefinition}, and the calculations of the ALP production spectra for photon coalescence and the Primakoff effect (with $ m_a >0 $) we can derive the `decay bound' of \cref{sec:decay} for an ALP with tree-level couplings to photons. This would amount to an update of \cite{Jaeckel:2017tud} (and an extension of section 5C in \cite{Caputo:2021rux} to larger masses).
	
	Recently, it has been pointed out that when the ALP decays quickly, the energy deposition in the progenitor star of low-energy core collapse supernovae can be large enough to affect their explosion kinematics \cite{Sung:2019xie,Caputo:2022mah}. A natural extension of our work would be to study how this additional effect can be used to constraint the ALP-electron coupling once the effective coupling to photons is taken into account.
	 
	Furthermore, more distant supernovae than SN1987A are expected to contribute to a diffuse background of ALPs \cite{Calore:2020tjw}. Such ALPs can convert into photons in astrophysical magnetic fields, which leads to an expected gamma-ray signal that can be searched for in data by e.g.~\textit{Fermi}-LAT \cite{Calore:2021hhn}. In an extension of the `decay bound' discussed in \cref{sec:decay} and the analysis in \cite{Calore:2020tjw}, it would be interesting to constrain the diffuse background of heavy ALPs through their loop-induced decays into photons.

    Moreover, we have restricted our analysis to the case of supernovae. This was justified by the fact that the effective coupling of ALPs to photons takes the largest values in hot environments where the temperature can overcome the electron mass. However, it would be interesting to understand if the effective coupling to photons can still play a role in other hot environments such as the core of red giants \cite{Carenza:2021osu} where temperatures can be as large as $10$ keV, or in the vicinity of low-luminosity black holes where the temperatures can also reach very large values \cite{Sadowski:2016fdh}. 
	
    In a broader context, the interplay between tree-level ALP-fermion couplings and loop-induced couplings to gauge fields is naturally not specific to electrons and photons. The case of light spin-0 particles coupled only to muons at tree-level was recently studied in \cite{Caputo:2021rux}.
    It was assumed that the effective ALP-photon coupling induced by a derivative coupling to muons vanishes for $ m_a \ll m_\mu \simeq 100 $~MeV, but as we have shown in \cref{sec:effectivePhotonCoupling} and discussed throughout this paper, this is not generally true and it would be interesting to re-consider the decay bound for large ALP masses in the light of our results.
	 
    We anticipate that the formalism developed and applied in this work will prove useful for future studies of astrophysical ALPs. Our work emphasises the important point that the effective couplings of ALPs are not arbitrary, but abide by quantum relations that make large hierarchies difficult to realise.
    
\acknowledgments{
    The authors would like to thank Tobias Fischer for sharing the simulation data with us, and in particular Pierluca Carenza for useful discussions about ALP physics and the supernovae bounds.
    RZF would also like to thank Cristina Manuel for insightful discussions on the use of the thermal masses. The work of RZF was supported by Spanish Ministry of Science and Innovation (PID2020-115845GB-I00/AEI/10.13039/501100011033), the Direcció General de Recerca del Departament d'Empresa i Coneixement (DGR) and by the EC through the program Marie Sklodowska-Curie COFUND (GA 801370)-Beatriu de Pinós. DM and EM are supported by the European Research Council under Grant No. 742104 and by the Swedish Research Council (VR) under grants 2018-03641 and 2019-02337.
}

\appendix

\section{Cross-sections for ALP production via electron Compton scattering and pair production}
\label{app: Cross-sections}

In this appendix we give the expressions for the cross-sections used in \cref{sec: Comparison with other bounds} to compute the ALP abundance contribution to $\Delta N_{\text{eff}}$ at neutrino decoupling.
\begin{itemize}[leftmargin=15pt]
    \item \textit{Pair production}: 
    \begin{equation*}
        \sigma_{e^-+e^+ \to a+ \gamma}= 4 \alpha_{\text{EM}}\,  \gae^2 \,\frac{m_e^2}{s}\, \frac{ \left(m_a^4-4 m_a^2 m_e^2+s^2\right) \tanh ^{-1}\left(\sqrt{1-\frac{4 m_e^2}{s}}\right)-m_a^2 \sqrt{s \left(s-4 m_e^2\right)}}{ \left(s-4 m_e^2\right) (s-m_a^2)  }
    \end{equation*}
    \item \textit{Compton scattering:}
     \begin{align*}
           \sigma_{e^\pm \gamma \to e^\pm a}=&\, 
           \alpha_{\text{EM}}\, \gae^2 \,\frac{m_e^2}{2s^2} \frac{ \sqrt{m_a^4-2 m_a^2 \left(m_e^2+s\right)+\left(m_e^2-s\right)^2}}{ \left(s-m_e^2\right)^3} \times \\
         &
        \left( m_a^2 \left(-m_e^4+2 m_e^2 s+7 s^2\right)+\left(m_e^2-s\right)^2 \left(m_e^2-3 s\right)- \right. \\
        & \left. \frac{4 s^2 \left(2 m_a^4-2 m_a^2 \left(m_e^2+s\right)+\left(m_e^2-s\right)^2\right) \coth ^{-1}\left(\frac{m_a^2-m_e^2-s}{\sqrt{m_a^4-2 m_a^2 \left(m_e^2+s\right)+\left(m_e^2-s\right)^2}}\right)}{\sqrt{m_a^4-2 m_a^2 \left(m_e^2+s\right)+\left(m_e^2-s\right)^2}}\right)
    \end{align*}
\end{itemize}
where we assumed $m_a<2m_e$ in both cases.

\bibliography{biblio}

\providecommand{\href}[2]{#2}\begingroup\raggedright\begin{thebibliography}{10}

\bibitem{Raffelt:1996wa}
G.~G. Raffelt, \emph{{Stars as laboratories for fundamental physics}: {The
  astrophysics of neutrinos, axions, and other weakly interacting particles}}.
\newblock University of Chicago Press, 5, 1996.

\bibitem{Peccei:1977hh}
R.~D. Peccei and H.~R. Quinn, \emph{{CP Conservation in the Presence of
  Instantons}},
  \href{http://dx.doi.org/10.1103/PhysRevLett.38.1440}{\emph{Phys. Rev. Lett.}
  {\bf 38} (1977) 1440--1443}.

\bibitem{Peccei:1977ur}
R.~D. Peccei and H.~R. Quinn, \emph{{Constraints Imposed by CP Conservation in
  the Presence of Instantons}},
  \href{http://dx.doi.org/10.1103/PhysRevD.16.1791}{\emph{Phys. Rev. D} {\bf
  16} (1977) 1791--1797}.

\bibitem{Weinberg:1977ma}
S.~Weinberg, \emph{{A New Light Boson?}},
  \href{http://dx.doi.org/10.1103/PhysRevLett.40.223}{\emph{Phys. Rev. Lett.}
  {\bf 40} (1978) 223--226}.

\bibitem{Wilczek:1977pj}
F.~Wilczek, \emph{{Problem of Strong $P$ and $T$ Invariance in the Presence of
  Instantons}}, \href{http://dx.doi.org/10.1103/PhysRevLett.40.279}{\emph{Phys.
  Rev. Lett.} {\bf 40} (1978) 279--282}.

\bibitem{Preskill:1982cy}
J.~Preskill, M.~B. Wise and F.~Wilczek, \emph{{Cosmology of the Invisible
  Axion}}, \href{http://dx.doi.org/10.1016/0370-2693(83)90637-8}{\emph{Phys.
  Lett. B} {\bf 120} (1983) 127--132}.

\bibitem{Abbott:1982af}
L.~F. Abbott and P.~Sikivie, \emph{{A Cosmological Bound on the Invisible
  Axion}}, \href{http://dx.doi.org/10.1016/0370-2693(83)90638-X}{\emph{Phys.
  Lett. B} {\bf 120} (1983) 133--136}.

\bibitem{Dine:1982ah}
M.~Dine and W.~Fischler, \emph{{The Not So Harmless Axion}},
  \href{http://dx.doi.org/10.1016/0370-2693(83)90639-1}{\emph{Phys. Lett. B}
  {\bf 120} (1983) 137--141}.

\bibitem{Svrcek:2006yi}
P.~Svrcek and E.~Witten, \emph{{Axions In String Theory}},
  \href{http://dx.doi.org/10.1088/1126-6708/2006/06/051}{\emph{JHEP} {\bf 06}
  (2006) 051}, [\href{https://arxiv.org/abs/hep-th/0605206}{{\tt
  hep-th/0605206}}].

\bibitem{Gelmini:1980re}
G.~B. Gelmini and M.~Roncadelli, \emph{{Left-Handed Neutrino Mass Scale and
  Spontaneously Broken Lepton Number}},
  \href{http://dx.doi.org/10.1016/0370-2693(81)90559-1}{\emph{Phys. Lett. B}
  {\bf 99} (1981) 411--415}.

\bibitem{Davidson:1981zd}
A.~Davidson and K.~C. Wali, \emph{{MINIMAL FLAVOR UNIFICATION VIA
  MULTIGENERATIONAL PECCEI-QUINN SYMMETRY}},
  \href{http://dx.doi.org/10.1103/PhysRevLett.48.11}{\emph{Phys. Rev. Lett.}
  {\bf 48} (1982) 11}.

\bibitem{Wilczek:1982rv}
F.~Wilczek, \emph{{Axions and Family Symmetry Breaking}},
  \href{http://dx.doi.org/10.1103/PhysRevLett.49.1549}{\emph{Phys. Rev. Lett.}
  {\bf 49} (1982) 1549--1552}.

\bibitem{Craig:2018kne}
N.~Craig, A.~Hook and S.~Kasko, \emph{{The Photophobic ALP}},
  \href{http://dx.doi.org/10.1007/JHEP09(2018)028}{\emph{JHEP} {\bf 09} (2018)
  028}, [\href{https://arxiv.org/abs/1805.06538}{{\tt 1805.06538}}].

\bibitem{Ferreira:2022egk}
R.~Z. Ferreira, M.~C.~D. Marsh and E.~M\"uller, \emph{{Do direct detection
  experiments constrain axionlike particles coupled to electrons?}},
  \href{https://arxiv.org/abs/2202.08858}{{\tt 2202.08858}}.

\bibitem{Ghosh:2020vti}
D.~Ghosh and D.~Sachdeva, \emph{{Constraints on Axion-Lepton coupling from Big
  Bang Nucleosynthesis}},
  \href{http://dx.doi.org/10.1088/1475-7516/2020/10/060}{\emph{JCAP} {\bf 10}
  (2020) 060}, [\href{https://arxiv.org/abs/2007.01873}{{\tt 2007.01873}}].

\bibitem{Ellis:1987pk}
J.~R. Ellis and K.~A. Olive, \emph{{Constraints on Light Particles From
  Supernova Sn1987a}},
  \href{http://dx.doi.org/10.1016/0370-2693(87)91710-2}{\emph{Phys. Lett. B}
  {\bf 193} (1987) 525}.

\bibitem{Raffelt:1987yt}
G.~Raffelt and D.~Seckel, \emph{{Bounds on Exotic Particle Interactions from SN
  1987a}}, \href{http://dx.doi.org/10.1103/PhysRevLett.60.1793}{\emph{Phys.
  Rev. Lett.} {\bf 60} (1988) 1793}.

\bibitem{Lucente:2021hbp}
G.~Lucente and P.~Carenza, \emph{{Supernova bound on axionlike particles
  coupled with electrons}},
  \href{http://dx.doi.org/10.1103/PhysRevD.104.103007}{\emph{Phys. Rev. D} {\bf
  104} (2021) 103007}, [\href{https://arxiv.org/abs/2107.12393}{{\tt
  2107.12393}}].

\bibitem{Fischer:2021jfm}
T.~Fischer, P.~Carenza, B.~Fore, M.~Giannotti, A.~Mirizzi and S.~Reddy,
  \emph{{Observable signatures of enhanced axion emission from protoneutron
  stars}}, \href{http://dx.doi.org/10.1103/PhysRevD.104.103012}{\emph{Phys.
  Rev. D} {\bf 104} (2021) 103012},
  [\href{https://arxiv.org/abs/2108.13726}{{\tt 2108.13726}}].

\bibitem{Depta:2020zbh}
P.~F. Depta, M.~Hufnagel and K.~Schmidt-Hoberg, \emph{{Updated BBN constraints
  on electromagnetic decays of MeV-scale particles}},
  \href{http://dx.doi.org/10.1088/1475-7516/2021/04/011}{\emph{JCAP} {\bf 04}
  (2021) 011}, [\href{https://arxiv.org/abs/2011.06519}{{\tt 2011.06519}}].

\bibitem{Carena:1988kr}
M.~Carena and R.~D. Peccei, \emph{{The Effective Lagrangian for Axion Emission
  From {SN1987A}}},
  \href{http://dx.doi.org/10.1103/PhysRevD.40.652}{\emph{Phys. Rev. D} {\bf 40}
  (1989) 652}.

\bibitem{Choi:1988xt}
K.~Choi, K.~Kang and J.~E. Kim, \emph{{Invisible Axion Emissions From
  {SN1987A}}}, \href{http://dx.doi.org/10.1103/PhysRevLett.62.849}{\emph{Phys.
  Rev. Lett.} {\bf 62} (1989) 849}.

\bibitem{Quevillon:2019zrd}
J.~Quevillon and C.~Smith, \emph{{Axions are blind to anomalies}},
  \href{http://dx.doi.org/10.1140/epjc/s10052-019-7304-4}{\emph{Eur. Phys. J.
  C} {\bf 79} (2019) 822}, [\href{https://arxiv.org/abs/1903.12559}{{\tt
  1903.12559}}].

\bibitem{MERTIG1991345}
R.~Mertig, M.~Böhm and A.~Denner, \emph{{Feyn Calc} - computer-algebraic
  calculation of feynman amplitudes},
  \href{http://dx.doi.org/https://doi.org/10.1016/0010-4655(91)90130-D}{\emph{Computer
  Physics Communications} {\bf 64} (1991) 345--359}.

\bibitem{Shtabovenko:2016sxi}
V.~Shtabovenko, R.~Mertig and F.~Orellana, \emph{{New Developments in FeynCalc
  9.0}}, \href{http://dx.doi.org/10.1016/j.cpc.2016.06.008}{\emph{Comput. Phys.
  Commun.} {\bf 207} (2016) 432--444},
  [\href{https://arxiv.org/abs/1601.01167}{{\tt 1601.01167}}].

\bibitem{Shtabovenko:2020gxv}
V.~Shtabovenko, R.~Mertig and F.~Orellana, \emph{{FeynCalc 9.3: New features
  and improvements}},
  \href{http://dx.doi.org/10.1016/j.cpc.2020.107478}{\emph{Comput. Phys.
  Commun.} {\bf 256} (2020) 107478},
  [\href{https://arxiv.org/abs/2001.04407}{{\tt 2001.04407}}].

\bibitem{Patel:2015tea}
H.~H. Patel, \emph{{Package-X: A Mathematica package for the analytic
  calculation of one-loop integrals}},
  \href{http://dx.doi.org/10.1016/j.cpc.2015.08.017}{\emph{Comput. Phys.
  Commun.} {\bf 197} (2015) 276--290},
  [\href{https://arxiv.org/abs/1503.01469}{{\tt 1503.01469}}].

\bibitem{Hahn:1998yk}
T.~Hahn and M.~Perez-Victoria, \emph{{Automatized one loop calculations in
  four-dimensions and D-dimensions}},
  \href{http://dx.doi.org/10.1016/S0010-4655(98)00173-8}{\emph{Comput. Phys.
  Commun.} {\bf 118} (1999) 153--165},
  [\href{https://arxiv.org/abs/hep-ph/9807565}{{\tt hep-ph/9807565}}].

\bibitem{Passarino:1978jh}
G.~Passarino and M.~J.~G. Veltman, \emph{{One Loop Corrections for e+ e-
  Annihilation Into mu+ mu- in the Weinberg Model}},
  \href{http://dx.doi.org/10.1016/0550-3213(79)90234-7}{\emph{Nucl. Phys. B}
  {\bf 160} (1979) 151--207}.

\bibitem{tHooft:1978jhc}
G.~'t~Hooft and M.~J.~G. Veltman, \emph{{Scalar One Loop Integrals}},
  \href{http://dx.doi.org/10.1016/0550-3213(79)90605-9}{\emph{Nucl. Phys. B}
  {\bf 153} (1979) 365--401}.

\bibitem{Breitenlohner:1977hr}
P.~Breitenlohner and D.~Maison, \emph{{Dimensional Renormalization and the
  Action Principle}}, \href{http://dx.doi.org/10.1007/BF01609069}{\emph{Commun.
  Math. Phys.} {\bf 52} (1977) 11--38}.

\bibitem{tHooft:1972tcz}
G.~'t~Hooft and M.~J.~G. Veltman, \emph{{Regularization and Renormalization of
  Gauge Fields}},
  \href{http://dx.doi.org/10.1016/0550-3213(72)90279-9}{\emph{Nucl. Phys. B}
  {\bf 44} (1972) 189--213}.

\bibitem{Bauer:2021mvw}
M.~Bauer, M.~Neubert, S.~Renner, M.~Schnubel and A.~Thamm, \emph{{Flavor probes
  of axion-like particles}},  \href{https://arxiv.org/abs/2110.10698}{{\tt
  2110.10698}}.

\bibitem{Chala:2020wvs}
M.~Chala, G.~Guedes, M.~Ramos and J.~Santiago, \emph{{Running in the ALPs}},
  \href{http://dx.doi.org/10.1140/epjc/s10052-021-08968-2}{\emph{Eur. Phys. J.
  C} {\bf 81} (2021) 181}, [\href{https://arxiv.org/abs/2012.09017}{{\tt
  2012.09017}}].

\bibitem{Bauer:2020jbp}
M.~Bauer, M.~Neubert, S.~Renner, M.~Schnubel and A.~Thamm, \emph{{The
  Low-Energy Effective Theory of Axions and ALPs}},
  \href{http://dx.doi.org/10.1007/JHEP04(2021)063}{\emph{JHEP} {\bf 04} (2021)
  063}, [\href{https://arxiv.org/abs/2012.12272}{{\tt 2012.12272}}].

\bibitem{Schwartz:2014sze}
M.~D. Schwartz, \emph{{Quantum Field Theory and the Standard Model}}.
\newblock Cambridge University Press, 3, 2014.

\bibitem{Weinberg:1996kr}
S.~Weinberg, \emph{{The quantum theory of fields. Vol. 2: Modern
  applications}}.
\newblock Cambridge University Press, 8, 2013.

\bibitem{Bauer:2017ris}
M.~Bauer, M.~Neubert and A.~Thamm, \emph{{Collider Probes of Axion-Like
  Particles}}, \href{http://dx.doi.org/10.1007/JHEP12(2017)044}{\emph{JHEP}
  {\bf 12} (2017) 044}, [\href{https://arxiv.org/abs/1708.00443}{{\tt
  1708.00443}}].

\bibitem{Calibbi:2020jvd}
L.~Calibbi, D.~Redigolo, R.~Ziegler and J.~Zupan, \emph{{Looking forward to
  Lepton-flavor-violating ALPs}},  \href{https://arxiv.org/abs/2006.04795}{{\tt
  2006.04795}}.

\bibitem{Caputo:2021rux}
A.~Caputo, G.~Raffelt and E.~Vitagliano, \emph{{Muonic boson limits: Supernova
  redux}}, \href{http://dx.doi.org/10.1103/PhysRevD.105.035022}{\emph{Phys.
  Rev. D} {\bf 105} (2022) 035022},
  [\href{https://arxiv.org/abs/2109.03244}{{\tt 2109.03244}}].

\bibitem{Payez:2014xsa}
A.~Payez, C.~Evoli, T.~Fischer, M.~Giannotti, A.~Mirizzi and A.~Ringwald,
  \emph{{Revisiting the SN1987A gamma-ray limit on ultralight axion-like
  particles}},
  \href{http://dx.doi.org/10.1088/1475-7516/2015/02/006}{\emph{JCAP} {\bf 02}
  (2015) 006}, [\href{https://arxiv.org/abs/1410.3747}{{\tt 1410.3747}}].

\bibitem{Lucente:2020whw}
G.~Lucente, P.~Carenza, T.~Fischer, M.~Giannotti and A.~Mirizzi, \emph{{Heavy
  axion-like particles and core-collapse supernovae: constraints and impact on
  the explosion mechanism}},
  \href{http://dx.doi.org/10.1088/1475-7516/2020/12/008}{\emph{JCAP} {\bf 12}
  (2020) 008}, [\href{https://arxiv.org/abs/2008.04918}{{\tt 2008.04918}}].

\bibitem{Ertas:2020xcc}
F.~Ertas and F.~Kahlhoefer, \emph{{On the interplay between astrophysical and
  laboratory probes of MeV-scale axion-like particles}},
  \href{http://dx.doi.org/10.1007/JHEP07(2020)050}{\emph{JHEP} {\bf 07} (2020)
  050}, [\href{https://arxiv.org/abs/2004.01193}{{\tt 2004.01193}}].

\bibitem{Brinkmann:1988vi}
R.~P. Brinkmann and M.~S. Turner, \emph{{Numerical Rates for Nucleon-Nucleon
  Axion Bremsstrahlung}},
  \href{http://dx.doi.org/10.1103/PhysRevD.38.2338}{\emph{Phys. Rev. D} {\bf
  38} (1988) 2338}.

\bibitem{Carenza:2019pxu}
P.~Carenza, T.~Fischer, M.~Giannotti, G.~Guo, G.~Mart\'\i{}nez-Pinedo and
  A.~Mirizzi, \emph{{Improved axion emissivity from a supernova via
  nucleon-nucleon bremsstrahlung}},
  \href{http://dx.doi.org/10.1088/1475-7516/2019/10/016}{\emph{JCAP} {\bf 10}
  (2019) 016}, [\href{https://arxiv.org/abs/1906.11844}{{\tt 1906.11844}}].

\bibitem{Bollig:2020xdr}
R.~Bollig, W.~DeRocco, P.~W. Graham and H.-T. Janka, \emph{{Muons in
  Supernovae: Implications for the Axion-Muon Coupling}},
  \href{http://dx.doi.org/10.1103/PhysRevLett.125.051104}{\emph{Phys. Rev.
  Lett.} {\bf 125} (2020) 051104},
  [\href{https://arxiv.org/abs/2005.07141}{{\tt 2005.07141}}].

\bibitem{Mezzacappa:1993gn}
A.~Mezzacappa and S.~W. Bruenn, \emph{{A numerical method for solving the
  neutrino Boltzmann equation coupled to spherically symmetric stellar core
  collapse}}, \href{http://dx.doi.org/10.1086/172395}{\emph{Astrophys. J.} {\bf
  405} (1993) 669--684}.

\bibitem{Liebendoerfer:2002xn}
M.~Liebendoerfer, O.~E.~B. Messer, A.~Mezzacappa, S.~W. Bruenn, C.~Y. Cardall
  and F.~K. Thielemann, \emph{{A Finite difference representation of neutrino
  radiation hydrodynamics for spherically symmetric general relativistic
  supernova simulations}},
  \href{http://dx.doi.org/10.1086/380191}{\emph{Astrophys. J. Suppl.} {\bf 150}
  (2004) 263--316}, [\href{https://arxiv.org/abs/astro-ph/0207036}{{\tt
  astro-ph/0207036}}].

\bibitem{Fischer:2013eka}
T.~Fischer, M.~Hempel, I.~Sagert, Y.~Suwa and J.~Schaffner-Bielich,
  \emph{{Symmetry energy impact in simulations of core-collapse supernovae}},
  \href{http://dx.doi.org/10.1140/epja/i2014-14046-5}{\emph{Eur. Phys. J. A}
  {\bf 50} (2014) 46}, [\href{https://arxiv.org/abs/1307.6190}{{\tt
  1307.6190}}].

\bibitem{Betranhandy:2022bvr}
A.~Betranhandy and E.~O'Connor, \emph{{Neutrino Driven Explosions aided by
  Axion Cooling in Multidimensional Simulations of Core-Collapse Supernovae}},
  \href{https://arxiv.org/abs/2204.00503}{{\tt 2204.00503}}.

\bibitem{Fischer:2016cyd}
T.~Fischer, S.~Chakraborty, M.~Giannotti, A.~Mirizzi, A.~Payez and A.~Ringwald,
  \emph{{Probing axions with the neutrino signal from the next galactic
  supernova}}, \href{http://dx.doi.org/10.1103/PhysRevD.94.085012}{\emph{Phys.
  Rev. D} {\bf 94} (2016) 085012},
  [\href{https://arxiv.org/abs/1605.08780}{{\tt 1605.08780}}].

\bibitem{Raffelt:1990yz}
G.~G. Raffelt, \emph{{Astrophysical methods to constrain axions and other novel
  particle phenomena}},
  \href{http://dx.doi.org/10.1016/0370-1573(90)90054-6}{\emph{Phys. Rept.} {\bf
  198} (1990) 1--113}.

\bibitem{Carenza:2021osu}
P.~Carenza and G.~Lucente, \emph{{Revisiting axion-electron bremsstrahlung
  emission rates in astrophysical environments}},
  \href{http://dx.doi.org/10.1103/PhysRevD.103.123024}{\emph{Phys. Rev. D} {\bf
  103} (2021) 123024}, [\href{https://arxiv.org/abs/2104.09524}{{\tt
  2104.09524}}].

\bibitem{Braaten:1992abc}
E.~{Braaten}, \emph{{Neutrino Emissivity of an Ultrarelativistic Plasma from
  Positron and Plasmino Annihilation}},
  \href{http://dx.doi.org/10.1086/171405}{\emph{apj} {\bf 392} (June, 1992)
  70}.

\bibitem{Raffelt:1985nk}
G.~G. Raffelt, \emph{{ASTROPHYSICAL AXION BOUNDS DIMINISHED BY SCREENING
  EFFECTS}}, \href{http://dx.doi.org/10.1103/PhysRevD.33.897}{\emph{Phys. Rev.
  D} {\bf 33} (1986) 897}.

\bibitem{Kopf:1997mv}
A.~Kopf and G.~Raffelt, \emph{{Photon dispersion in a supernova core}},
  \href{http://dx.doi.org/10.1103/PhysRevD.57.3235}{\emph{Phys. Rev. D} {\bf
  57} (1998) 3235--3241}, [\href{https://arxiv.org/abs/astro-ph/9711196}{{\tt
  astro-ph/9711196}}].

\bibitem{Caputo:2022mah}
A.~Caputo, H.-T. Janka, G.~Raffelt and E.~Vitagliano, \emph{{Low-Energy
  Supernovae Severely Constrain Radiative Particle Decays}},
  \href{http://dx.doi.org/10.1103/PhysRevLett.128.221103}{\emph{Phys. Rev.
  Lett.} {\bf 128} (2022) 221103},
  [\href{https://arxiv.org/abs/2201.09890}{{\tt 2201.09890}}].

\bibitem{Chang:2016ntp}
J.~H. Chang, R.~Essig and S.~D. McDermott, \emph{{Revisiting Supernova 1987A
  Constraints on Dark Photons}},
  \href{http://dx.doi.org/10.1007/JHEP01(2017)107}{\emph{JHEP} {\bf 01} (2017)
  107}, [\href{https://arxiv.org/abs/1611.03864}{{\tt 1611.03864}}].

\bibitem{Chang:2018rso}
J.~H. Chang, R.~Essig and S.~D. McDermott, \emph{{Supernova 1987A Constraints
  on Sub-GeV Dark Sectors, Millicharged Particles, the QCD Axion, and an
  Axion-like Particle}},
  \href{http://dx.doi.org/10.1007/JHEP09(2018)051}{\emph{JHEP} {\bf 09} (2018)
  051}, [\href{https://arxiv.org/abs/1803.00993}{{\tt 1803.00993}}].

\bibitem{Caputo:2022rca}
A.~Caputo, G.~Raffelt and E.~Vitagliano, \emph{{Radiative transfer in stars by
  feebly interacting bosons}},  \href{https://arxiv.org/abs/2204.11862}{{\tt
  2204.11862}}.

\bibitem{Chupp:1989kx}
E.~L. Chupp, W.~T. Vestrand and C.~Reppin, \emph{{Experimental Limits on the
  Radiative Decay of {SN1987A} Neutrinos}},
  \href{http://dx.doi.org/10.1103/PhysRevLett.62.505}{\emph{Phys. Rev. Lett.}
  {\bf 62} (1989) 505--508}.

\bibitem{Jaeckel:2017tud}
J.~Jaeckel, P.~C. Malta and J.~Redondo, \emph{{Decay photons from the axionlike
  particles burst of type II supernovae}},
  \href{http://dx.doi.org/10.1103/PhysRevD.98.055032}{\emph{Phys. Rev. D} {\bf
  98} (2018) 055032}, [\href{https://arxiv.org/abs/1702.02964}{{\tt
  1702.02964}}].

\bibitem{Jaffe:1995sw}
A.~H. Jaffe and M.~S. Turner, \emph{{Gamma-rays and the decay of neutrinos from
  SN1987A}}, \href{http://dx.doi.org/10.1103/PhysRevD.55.7951}{\emph{Phys. Rev.
  D} {\bf 55} (1997) 7951--7959},
  [\href{https://arxiv.org/abs/astro-ph/9601104}{{\tt astro-ph/9601104}}].

\bibitem{Oberauer:1993yr}
L.~Oberauer, C.~Hagner, G.~Raffelt and E.~Rieger, \emph{{Supernova bounds on
  neutrino radiative decays}},
  \href{http://dx.doi.org/10.1016/0927-6505(93)90004-W}{\emph{Astropart. Phys.}
  {\bf 1} (1993) 377--386}.

\bibitem{VanTilburg:2020jvl}
K.~Van~Tilburg, \emph{{Stellar basins of gravitationally bound particles}},
  \href{http://dx.doi.org/10.1103/PhysRevD.104.023019}{\emph{Phys. Rev. D} {\bf
  104} (2021) 023019}, [\href{https://arxiv.org/abs/2006.12431}{{\tt
  2006.12431}}].

\bibitem{Aguilar-Arevalo:2021wjq}
A.~A. Aguilar-Arevalo et~al., \emph{{Axion-Like Particles at Coherent
  CAPTAIN-Mills}},  \href{https://arxiv.org/abs/2112.09979}{{\tt 2112.09979}}.

\bibitem{Darme:2020sjf}
L.~Darm\'e, F.~Giacchino, E.~Nardi and M.~Raggi, \emph{{Invisible decays of
  axion-like particles: constraints and prospects}},
  \href{http://dx.doi.org/10.1007/JHEP06(2021)009}{\emph{JHEP} {\bf 06} (2021)
  009}, [\href{https://arxiv.org/abs/2012.07894}{{\tt 2012.07894}}].

\bibitem{NA64:2021aiq}
{\scshape NA64} collaboration, Y.~M. Andreev et~al., \emph{{Search for
  pseudoscalar bosons decaying into $e^+e^-$ pairs in the NA64 experiment at
  the CERN SPS}},
  \href{http://dx.doi.org/10.1103/PhysRevD.104.L111102}{\emph{Phys. Rev. D}
  {\bf 104} (2021) L111102}, [\href{https://arxiv.org/abs/2104.13342}{{\tt
  2104.13342}}].

\bibitem{Morel:2020dww}
L.~Morel, Z.~Yao, P.~Clad\'e and S.~Guellati-Kh\'elifa, \emph{{Determination of
  the fine-structure constant with an accuracy of 81 parts per trillion}},
  \href{http://dx.doi.org/10.1038/s41586-020-2964-7}{\emph{Nature} {\bf 588}
  (2020) 61--65}.

\bibitem{Essig:2010gu}
R.~Essig, R.~Harnik, J.~Kaplan and N.~Toro, \emph{{Discovering New Light States
  at Neutrino Experiments}},
  \href{http://dx.doi.org/10.1103/PhysRevD.82.113008}{\emph{Phys. Rev. D} {\bf
  82} (2010) 113008}, [\href{https://arxiv.org/abs/1008.0636}{{\tt
  1008.0636}}].

\bibitem{XENON:2019gfn}
{\scshape XENON} collaboration, E.~Aprile et~al., \emph{{Light Dark Matter
  Search with Ionization Signals in XENON1T}},
  \href{http://dx.doi.org/10.1103/PhysRevLett.123.251801}{\emph{Phys. Rev.
  Lett.} {\bf 123} (2019) 251801},
  [\href{https://arxiv.org/abs/1907.11485}{{\tt 1907.11485}}].

\bibitem{XENON:2020rca}
{\scshape XENON} collaboration, E.~Aprile et~al., \emph{{Excess electronic
  recoil events in XENON1T}},
  \href{http://dx.doi.org/10.1103/PhysRevD.102.072004}{\emph{Phys. Rev. D} {\bf
  102} (2020) 072004}, [\href{https://arxiv.org/abs/2006.09721}{{\tt
  2006.09721}}].

\bibitem{Bechis:1979kp}
D.~J. Bechis, T.~W. Dombeck, R.~W. Ellsworth, E.~V. Sager, P.~H. Steinberg,
  L.~J. Teig et~al., \emph{{Search for Axion Production in Low-energy Electron
  Bremsstrahlung}},
  \href{http://dx.doi.org/10.1103/PhysRevLett.42.1511}{\emph{Phys. Rev. Lett.}
  {\bf 42} (1979) 1511}.

\bibitem{BaBar:2017tiz}
{\scshape BaBar} collaboration, J.~P. Lees et~al., \emph{{Search for Invisible
  Decays of a Dark Photon Produced in ${e}^{+}{e}^{-}$ Collisions at BaBar}},
  \href{http://dx.doi.org/10.1103/PhysRevLett.119.131804}{\emph{Phys. Rev.
  Lett.} {\bf 119} (2017) 131804},
  [\href{https://arxiv.org/abs/1702.03327}{{\tt 1702.03327}}].

\bibitem{Bjorken:1988as}
J.~D. Bjorken, S.~Ecklund, W.~R. Nelson, A.~Abashian, C.~Church, B.~Lu et~al.,
  \emph{{Search for Neutral Metastable Penetrating Particles Produced in the
  SLAC Beam Dump}},
  \href{http://dx.doi.org/10.1103/PhysRevD.38.3375}{\emph{Phys. Rev. D} {\bf
  38} (1988) 3375}.

\bibitem{Fields:2019pfx}
B.~D. Fields, K.~A. Olive, T.-H. Yeh and C.~Young, \emph{{Big-Bang
  Nucleosynthesis after Planck}},
  \href{http://dx.doi.org/10.1088/1475-7516/2020/03/010}{\emph{JCAP} {\bf 03}
  (2020) 010}, [\href{https://arxiv.org/abs/1912.01132}{{\tt 1912.01132}}].

\bibitem{Ferreira:2018vjj}
R.~Z. Ferreira and A.~Notari, \emph{{Observable Windows for the QCD Axion
  Through the Number of Relativistic Species}},
  \href{http://dx.doi.org/10.1103/PhysRevLett.120.191301}{\emph{Phys. Rev.
  Lett.} {\bf 120} (2018) 191301},
  [\href{https://arxiv.org/abs/1801.06090}{{\tt 1801.06090}}].

\bibitem{DEramo:2018vss}
F.~D'Eramo, R.~Z. Ferreira, A.~Notari and J.~L. Bernal, \emph{{Hot Axions and
  the $H_0$ tension}},
  \href{http://dx.doi.org/10.1088/1475-7516/2018/11/014}{\emph{JCAP} {\bf 11}
  (2018) 014}, [\href{https://arxiv.org/abs/1808.07430}{{\tt 1808.07430}}].

\bibitem{Kawasaki:2017bqm}
M.~Kawasaki, K.~Kohri, T.~Moroi and Y.~Takaesu, \emph{{Revisiting Big-Bang
  Nucleosynthesis Constraints on Long-Lived Decaying Particles}},
  \href{http://dx.doi.org/10.1103/PhysRevD.97.023502}{\emph{Phys. Rev. D} {\bf
  97} (2018) 023502}, [\href{https://arxiv.org/abs/1709.01211}{{\tt
  1709.01211}}].

\bibitem{Cadamuro:2011fd}
D.~Cadamuro and J.~Redondo, \emph{{Cosmological bounds on pseudo
  Nambu-Goldstone bosons}},
  \href{http://dx.doi.org/10.1088/1475-7516/2012/02/032}{\emph{JCAP} {\bf 02}
  (2012) 032}, [\href{https://arxiv.org/abs/1110.2895}{{\tt 1110.2895}}].

\bibitem{Sung:2019xie}
A.~Sung, H.~Tu and M.-R. Wu, \emph{{New constraint from supernova explosions on
  light particles beyond the Standard Model}},
  \href{http://dx.doi.org/10.1103/PhysRevD.99.121305}{\emph{Phys. Rev. D} {\bf
  99} (2019) 121305}, [\href{https://arxiv.org/abs/1903.07923}{{\tt
  1903.07923}}].

\bibitem{Calore:2020tjw}
F.~Calore, P.~Carenza, M.~Giannotti, J.~Jaeckel and A.~Mirizzi, \emph{{Bounds
  on axionlike particles from the diffuse supernova flux}},
  \href{http://dx.doi.org/10.1103/PhysRevD.102.123005}{\emph{Phys. Rev. D} {\bf
  102} (2020) 123005}, [\href{https://arxiv.org/abs/2008.11741}{{\tt
  2008.11741}}].

\bibitem{Calore:2021hhn}
F.~Calore, P.~Carenza, C.~Eckner, T.~Fischer, M.~Giannotti, J.~Jaeckel et~al.,
  \emph{{3D template-based $Fermi$-LAT constraints on the diffuse supernova
  axion-like particle background}},
  \href{https://arxiv.org/abs/2110.03679}{{\tt 2110.03679}}.

\bibitem{Sadowski:2016fdh}
A.~Sadowski, M.~Wielgus, R.~Narayan, D.~Abarca, J.~C. McKinney and A.~Chael,
  \emph{{Radiative, two-temperature simulations of low luminosity black hole
  accretion flows in general relativity}},
  \href{http://dx.doi.org/10.1093/mnras/stw3116}{\emph{Mon. Not. Roy. Astron.
  Soc.} {\bf 466} (2017) 705--725},
  [\href{https://arxiv.org/abs/1605.03184}{{\tt 1605.03184}}].

\end{thebibliography}\endgroup

\end{document}